\documentclass[12pt,notitlepage]{amsart}%
\usepackage{amssymb}
\usepackage{amsfonts}
\usepackage{graphicx}
\usepackage{amscd}
\usepackage{graphicx}
\usepackage{amsmath}
\usepackage{amsxtra}
\usepackage{footmisc}
\usepackage{acronym}%
\theoremstyle{plain}

\newtheorem{remark}{Remark}

\numberwithin{equation}{section}
\numberwithin{theorem}{section}
\numberwithin{lemma}{section}
\numberwithin{proposition}{section}
\numberwithin{corollary}{section}

\ifx\pdfoutput\relax\let\pdfoutput=\undefined\fi
\newcount\msipdfoutput
\ifx\pdfoutput\undefined\else
\ifcase\pdfoutput\else
\ifx\paperwidth\undefined\else
\ifdim\paperheight=0pt\relax\else\pdfpageheight\paperheight\fi
\ifdim\paperwidth=0pt\relax\else\pdfpagewidth\paperwidth\fi
\fi\fi\fi
\begin{document}
\title[The Wigner's Friend Paradox]{Wavefunctions localization, and the Wigner's Friend Paradox in a Framework of
Discrete-Space Hypothesis}
\author[Z\'{u}\~{n}iga-Galindo]{W. A. Z\'{u}\~{n}iga-Galindo}
\address{University of Texas Rio Grande Valley\\
School of Mathematical \& Statistical Sciences\\
One West University Blvd\\
Brownsville, TX 78520, United States}
\email{wilson.zunigagalindo@utrgv.edu}

\begin{abstract}
We present a resolution of the Wigner's Friend paradox within a framework of
quantum mechanics (QM) on the space $\mathbb{R}\times\mathbb{Q}_{p}$, where
$\mathbb{Q}_{p}$ denotes the field of $p$-adic numbers, regarded as a model of
discrete microscopic space at the Planck--Bronstein scale, and $\mathbb{R}$ is
a model of the macroscopic space. Time is treated as a real variable, so the
model is framed in the Dirac-von Neumann formalism. In this framework,
wavefunction collapse does not require a postulate about observers. Here,
collapse means spatial localization (localization in position), described by
the standard projection rule of the Dirac--von Neumann formalism; the
framework specifies when this rule applies and which projection is involved.
This mechanism resembles the one proposed by Ghirardi-Rimini-Weber, but in our
framework the region of localization is determined by the space geometry.
Between measurements, systems evolve according to the Schr\"{o}dinger equation
with non-local Hamiltonians. During a measurement interaction, the apparatus
probes the microscopic space through a continuous map from $\mathbb{Q}_{p}$ to
$\mathbb{R}$, and the wavefunction localizes on a ball of $\mathbb{Q}_{p}$
selected by the Born rule. We model both Wigner and his Friend as apparatuses
and show that each produces a definite reading through independent
applications of this localization mechanism, without observers or information
exchange; the readings persist as records, which Wigner cannot undo. We discuss the principal no-go theorems for extended Wigner's Friend
scenarios --- Frauchiger--Renner, Brukner, Bong et al., Gu\'{e}rin et al., and
Ormrod--Barrett --- theorem by theorem, and identify, in each case, the premise
that fails in the framework or has no counterpart in it. When the Friends are quantum systems, as in the photonic
experiments, the framework reproduces the quantum correlations and the
observed violation of the Local Friendliness inequalities; when the Friends
are apparatuses, their measurements cannot be reversed, and the framework
predicts that such violations cease. In an appendix, we
illustrate the collapse mechanism through a toy model of a particle in a box. The
framework is non-local, and the apparatus readings are absolute: they do not
depend on any observer.

\end{abstract}
\maketitle

\section{Introduction}

The measurement problem is one of the oldest and most resilient difficulties
in the foundations of quantum mechanics~(QM). In the standard formulation of
QM, the postulate of wavefunction collapse is introduced as a separate rule
that interrupts unitary Schr\"{o}dinger evolution whenever a measurement is
performed. This postulate has long been regarded as unsatisfactory because it
is vague (what counts as a measurement? what counts as an observer?),
non-dynamical (collapse is instantaneous and non-unitary), and in tension with
the universal applicability of the Schr\"{o}dinger equation.

The Wigner's Friend thought experiment~\cite{Wigner61} vividly conveys the
tension. Wigner imagines a friend~$F$ who performs a quantum measurement on a
spin-$\tfrac{1}{2}$ particle $S$ inside a perfectly isolated laboratory.
According to standard QM, after the measurement, $F$ has a definite result.
Yet Wigner, who remains outside the sealed laboratory and has not observed any
outcome, is entitled to apply unitary evolution to the entire closed system
$(S+\text{apparatus}+F+\text{lab})$. He therefore assigns an entangled
superposition to the joint state. The conflict is stark: $F$ claims a definite
fact; $W$ assigns a superposition. Neither description can be
straightforwardly dismissed within the standard formalism. The paradox has
generated an extensive literature proposing resolutions ranging from
many-worlds interpretations~\cite{Everett57} to relational QM~\cite{Rovelli96}%
, consistent histories~\cite{Griffiths02}, and QBism~\cite{FuchsMermin14}.

In \cite{Zuniga-2026}, the author proposed a new mechanism for wavefunction
collapse rooted in a structural hypothesis about the nature of space itself.
The\textrm{ space discreteness hypothesis} asserts, following the Bronstein
inequality, \cite{Bronstein}-\cite{Garay95}, that space at short distances is
radically different from its large-scale, manifold-like appearance.
Specifically, \cite{Zuniga-2026} models physical space as $(\mathbb{R}%
\times\mathcal{X})^{3}$, where $\mathcal{X}$\ is a totally disconnected
topological space. For simplicity, we work with a `one-dimensional model'
$\mathbb{R}\times\mathcal{X}$. In this framework, QM\ on the Hilbert space
$L^{2}(\mathbb{R}\times\mathcal{X})$ becomes a natural extension of QM\ on
$L^{2}(\mathbb{R})$. In sectors of type $\mathbb{R}\times\left\{
\alpha\right\}  \simeq\mathbb{R}$, with $\alpha\in\mathcal{X}$, the QM is
reduced to the standard case, and relativity is valid. While in sectors of
type $\left\{  \beta\right\}  \times\mathcal{X}\simeq\mathcal{X}$, with
$\beta\in\mathbb{R}$, relativity is not valid, and QM\ on $L^{2}(\mathcal{X})$
is a non-local theory.

In \cite{Zuniga-2026}, $\mathcal{X}$ is a model of the microscopic space,
while $\mathbb{R}$ models the macroscopic one.\ The image of $\mathcal{X}$ by
any\ continuous map $\mathcal{M}:\mathcal{X}\rightarrow\mathbb{R}$ gives a
distorted copy of the microscopic space into the macroscopic one. In QM\ on
$L^{2}(\mathbb{R}\times\mathcal{X})$, the Schr\"{o}dinger equation governs the
evolution of a system between measurements, and a measurement is described as
an interaction between a microscopic system and an apparatus, mediated by a
continuous map $\mathcal{M}:\mathcal{X}\rightarrow\mathbb{R}$. During this
interaction the wavefunction localizes (collapses) in space, which produces
definite readings in the measurement apparatuses. The localization is
described by the standard projection rule of QM. What the framework adds is a
physical specification of what counts as a measurement, namely an interaction
mediated by the map $\mathcal{M}$, and of the projection involved, which is
determined by the geometry of $\mathbb{R}\times\mathcal{X}$ and by
$\mathcal{M}$. This does not need intelligent agents (observers), and no
postulate about observers is required. The Born rule is retained: it
determines where the localization occurs.

It is well known that an objective collapse mechanism removes the conflict
between the descriptions of Wigner and his Friend; see, e.g., the discussion
of collapse theories in \cite{Frauchiger-Renner}. The present paper aims to
show that an extension of the framework of \cite{Zuniga-2026} provides a
self-consistent and mathematically precise resolution of the original Wigner's
Friend paradox, in which Wigner and his Friend are just apparatuses producing
definite readings without information exchange, and to identify what is
specific to this resolution: (i) the localization is not a stochastic process
and requires neither new physical constants nor a modification of the
Schr\"{o}dinger equation; (ii) Wigner's measurement, unlike his Friend's,
requires the microscopic space $\mathbb{R}\times\mathbb{Q}_{p}$ rather than
$\mathbb{Q}_{p}$; (iii) the boundary between quantum systems and apparatuses
is given by a physical criterion: a system acts as an apparatus when it
interacts with the quantum system through the map $\mathcal{M}$; and (iv) this
criterion leads to a prediction for extended Wigner's Friend experiments with
Friends of increasing size.

The paper is organized as follows. In Section \ref{Section_1}, we review the
space discreteness hypothesis and its mathematical implementation using
$p$-adic numbers ($\mathcal{X}=\mathbb{Q}_{p}$). A central problem was to know
if QM on $L^{2}(\mathbb{Q}_{p})$ describes physical systems. In
\cite{Zuniga-2026}, \cite{Zuniga-QM-2}, the author established that QM on
$\mathbb{C}^{N}$ can be recast as QM\ on $L^{2}(\mathbb{Z}_{p})$, \ where
$\mathbb{Z}_{p}$ is the unit ball in $\mathbb{Q}_{p}$. This means that any
unitary operator of type $e^{-it\left[  H_{i,j}\right]  }$, where $\left[
H_{i,j}\right]  $ is a Hermitian matrix, admits a continuous extension
$e^{-it\boldsymbol{H}}:L^{2}(\mathbb{Z}_{p})\rightarrow L^{2}(\mathbb{Z}_{p}%
)$, where $\boldsymbol{H}$ is a self-adjoint operator.

Section \ref{Section_2} reviews and expands the collapse mechanism for
wavefunctions introduced in \cite{Zuniga-2026}. In Subsection
\ref{Section_Meas_I}, we review the collapse mechanism using $\mathbb{Q}_{p}$
as a model of microscopic space; the technical details are given in the
appendices. This is the framework needed to explain the measurement performed
by Wigner's Friend. We show that the update rule for the state during a
measurement is the standard projection rule of QM, and state a criterion that
distinguishes apparatuses from quantum systems. In Subsection
\ref{Section_Meas_II}, we discuss the measurement problem using $\mathbb{R}%
\times\mathbb{Q}_{p}$ as a model of the microscopic space. This is the
framework we need to explain Wigner's measurement. In Subsection
\ref{Section_Further}, we compare the mechanism with the
Ghirardi--Rimini--Weber (GRW) and continuous spontaneous localization (CSL)
theories and with Bohmian mechanics, and discuss the alternative reading in
which the localization is only a projection of a global state.

In Section \ref{Section_4}, we use the localization mechanism to explain the
Wigner's Friend paradox, and show that the readings persist as records that
cannot be undone. In Section \ref{Section_5}, we review the no-go
theorems associated with extended Wigner's Friend scenarios, grouped according
to the notion on which they rely: agents, records of observed events, and the
Local Friendliness inequalities. We first discuss which systems act as
apparatuses in our framework. Since the framework is non-local, the Local
Friendliness theorem does not constrain it; we show that the framework
reproduces the quantum correlations of the Local Friendliness scenario when
the Friends are quantum systems (Appendix E), and describe the experimental consequences of
the framework. Finally, in Section \ref{Section_6}, we present the
conclusions. The basic mathematical results needed are collected in Appendices
A--C. In Appendix D, we discuss a toy model for the measurement of the energy
levels of a particle in a box, comparing the standard description based on QM
on $L^{2}(\mathbb{R})$, \cite{Norsen}, with the description based on QM on
$L^{2}(\mathbb{R}\times\mathbb{Q}_{p})$. In Appendix F, we analyze the
alternative reading in which the localization is only a projection, and show
that the update rule is what guarantees the persistence of records.

\section{\label{Section_1}QM in the Discrete-Space Framework}

\subsection{The Space Discreteness Hypothesis}

The Bronstein inequality \cite{Bronstein}-\cite{Garay95} arises when one
combines the Heisenberg uncertainty principle with the requirement that a
spatial measurement does not cause gravitational collapse. It implies a
fundamental minimal length scale $\ell_{B}$, below which the notion of a
smooth, locally Euclidean spatial manifold breaks down; the space becomes
discrete at very short distances. In \cite{Zuniga-2026}, the space at the
Planck--Bronstein scale is modeled by a \textrm{totally disconnected
topological space} $\mathcal{X}$. By definition, the connected components of
$\mathcal{X}$ are singletons. This implies that there is no continuous curve
$\gamma:[0,1]\rightarrow\mathcal{X}$ with $\gamma(0)=x\neq\gamma(1)=y$; in
particular, worldlines in the relativistic sense do not exist in $\mathcal{X}$.

In principle, $\mathcal{X}$ is a model at the Planck--Bronstein scale (very
high-energy regimes); in \cite{Zuniga-QM-2}, the author showed that
continuous-time quantum walks can be obtained using the space-times of type
$\mathbb{R}\times\mathcal{X}$. For this reason, the author has proposed using
$\mathbb{R}\times\mathcal{X}$ as a model of the microscopic space-time. It is
well-known that $\mathbb{R}\times\mathbb{R}^{3}$ is a very good model of the
space-time at large distances. Since $\mathbb{R}^{3}$ (or in general,
$\mathbb{R}^{N}$) cannot contain a homeomorphic copy of $\mathcal{X}$, it is
natural to propose $\mathbb{R}\times(\mathbb{R}\times\mathcal{X})^{3}$ as a
model of the space-time. For simplicity, we work with a single spatial
variable, so the new model is $\mathbb{R}\times(\mathbb{R}\times\mathcal{X})$.
This space-time contains sectors of type $\mathbb{R}\times(\mathbb{R}%
\times\left\{  \alpha\right\}  )$, $\alpha\in\mathcal{X}$, where relativity
provides excellent models, some of which are compatible with QM. But it also
contains sectors of type $\mathbb{R}\times(\left\{  \beta\right\}
\times\mathcal{X})$, $\beta\in\mathbb{R}$, where space is totally
disconnected, and the QM on such space-times is incompatible with relativity.

\subsection{The $p$-adic space}

From now on, we use $p$ to denote a fixed prime number. The field of $p$-adic
numbers $\mathbb{Q}_{p}$ is a paradigmatic example of totally disconnected,
locally compact, space that carries a natural Haar measure.\ Any non-zero
$p$-adic number $x$ has a unique expansion of the form%
\begin{equation}
x=x_{-k}p^{-k}+x_{-k+1}p^{-k+1}+\ldots+x_{0}+x_{1}p+\ldots,\text{ }
\label{p-adic-number}%
\end{equation}
with $x_{-k}\neq0$, where $k$ is an integer, and the $x_{j}$s\ are numbers
from the set $\left\{  0,1,\ldots,p-1\right\}  $. The set of all possible
sequences of the form (\ref{p-adic-number}) constitutes the field of $p$-adic
numbers $\mathbb{Q}_{p}$. There are natural field operations, sum and
multiplication, on series of form (\ref{p-adic-number}). There is also a norm
in $\mathbb{Q}_{p}$ defined as $\left\vert x\right\vert _{p}=p^{-ord(x)}$,
where $ord_{p}(x)=ord(x)=-k$, for a nonzero $p$-adic number $x$. By definition
$ord(0)=\infty$. The field of $p$-adic numbers with the distance induced by
$\left\vert \cdot\right\vert _{p}$ is a complete ultrametric space. The
ultrametric property refers to the fact that $\left\vert x-y\right\vert
_{p}\leq\max\left\{  \left\vert x-z\right\vert _{p},\left\vert z-y\right\vert
_{p}\right\}  $ for any $x$, $y$, $z$ in $\mathbb{Q}_{p}$. The $p$-adic
integers, which are sequences of the form (\ref{p-adic-number}) with $-k\geq
0$, constitute the unit ball $\mathbb{Z}_{p}$. The unit ball is an infinite
rooted tree with fractal structure. As a topological space $\mathbb{Q}_{p}%
$\ is homeomorphic to a Cantor-like subset of the real line. There is a
natural integration theory so that $\int_{\mathbb{Q}_{p}}\varphi\left(
x\right)  dx$ gives a well-defined complex number. The measure $dx$ is the
Haar measure of $\mathbb{Q}_{p}$. For an in-depth discussion, the reader may
consult \cite{V-V-Z}-\cite{Zuniga-Textbook}. In Appendix A, we give a quick
summary of the essential aspects of $p$-adic analysis required here.

Here we take $\mathcal{X}=\mathbb{Q}_{p}$ as in \cite{Zuniga-2026},
\cite{Zuniga-QM-2}. There are several reasons for this choice. First, in the
1980s, Volovich conjectured that the space has a $p$-adic nature at the Planck
scale, \cite{Volovich}-\cite{Varadarajan}. Second, $p$-adic QM has a physical
meaning: it describes the continuous-time random walks on graphs used in
quantum computing, \cite{Zuniga-QM-2}, \cite{Zuniga-Mayes}-\cite{Zuniga-CTQW},
and the $p$-adic Jackiw-Rebbi model, \cite{Zuniga-JMP-2026}. Furthermore, QM
on $\mathbb{C}^{N}$ can be recast as QM on $L^{2}(\mathbb{Z}_{p})$, as we
discuss below. Now, QM on $\mathbb{C}^{N}$ is a relevant tool in quantum
computing and in the formulation of the Bell inequalities, among several
applications. We argue that these arguments provide strong support for the
choice $\mathcal{X}=\mathbb{Q}_{p}$. However, at this moment, we do not know
whether Volovich's conjecture is true. The author conjectures that the
collapse mechanism introduced in \cite{Zuniga-2026} works for any totally
disconnected space $\mathcal{X}$. The selection of a particular state space of
type $L^{2}\left(  \ \mathbb{R}\times\mathcal{X}\right)  $ depends on the
quantum model to be constructed, and its ultimate validation should come from
comparison with experiments.

\subsection{QM on $L^{2}(\mathbb{R})%
{\textstyle\bigotimes}
L^{2}(\mathbb{Q}_{p})$}

\cite{Zuniga-2026} works in the Dirac--von Neumann formalism with Hilbert
space
\[
\mathcal{H}=L^{2}(\mathbb{R})%
{\textstyle\bigotimes}
L^{2}(\mathbb{Q}_{p})=L^{2}(\mathbb{R}\times\mathbb{Q}_{p},\,dx_{\infty}%
dx_{p}),
\]
where $dx_{\infty}$ is Lebesgue measure on $\mathbb{R}$ and $dx_{p}$ is Haar
measure on $\mathbb{Q}_{p}$, and $dx_{\infty}dx_{p}$\ denotes their product.
We use $x_{\infty}$ as a coordinate for $\mathbb{R}$, and $x_{p}$\ as
a\ coordinate for\ $\mathbb{Q}_{p}$. This notation has a number-theoretic
motivation. A state is a unit vector $\psi\in\mathcal{H}$. Observables are
self-adjoint operators on $\mathcal{H}$. Time evolution is governed by the
Schr\"{o}dinger equation
\begin{equation}
i\,\partial_{t}\,\psi(t,x_{\infty},x_{p})\;=\;\boldsymbol{H}\,\psi
(t,x_{\infty},x_{p}),\qquad(t,x_{\infty},x_{p})\in\mathbb{R}\times
\mathbb{R}\times\mathbb{Q}_{p}. \label{eq:schrodinger}%
\end{equation}
In \cite{Zuniga-2026}, it has been proposed using Hamiltonians of the form%
\[
\boldsymbol{H=H}_{\infty}%
{\textstyle\bigotimes}
1_{\mathbb{Q}_{p}}+\boldsymbol{P}(\boldsymbol{H}_{\infty},\boldsymbol{H}%
_{p})+1_{\mathbb{R}}%
{\textstyle\bigotimes}
\boldsymbol{H}_{p},
\]
where $\boldsymbol{H}_{\infty}$ is a self-adjoint operator with dense domain
in $L^{2}(\mathbb{R})$, $\boldsymbol{H}_{p}$ is a self-adjoint operator with
dense domain in $L^{2}(\mathbb{Q}_{p})$, $1_{\mathbb{Q}_{p}}$, $1_{\mathbb{R}%
}$ are the identity operators, and $\boldsymbol{P}(\boldsymbol{H}_{\infty
},\boldsymbol{H}_{p})$ denotes a polynomial operator without constant term.

The $p$-adic free Schr\"{o}dinger equation%
\[
\,\partial_{t}\,\psi_{p}(t,x_{p})\;=\;\boldsymbol{H}_{p}^{\text{free}}%
\,\psi_{p}(t,x_{p}),\qquad(t,x_{p})\in\mathbb{R}\times\mathbb{Q}_{p},
\]
is obtained from a $p$-adic heat equation, by the Wick rotation,%
\begin{equation}
\,\partial_{t}\,u(t,x_{p})\;=\;-\boldsymbol{H}_{p}^{\text{free}}%
\,u(t,x_{p}),\qquad(t,x_{p})\in\mathbb{R}\times\mathbb{Q}_{p}.
\label{p_adic_Heat_eq}%
\end{equation}
The term `heat equation,' means that (\ref{p_adic_Heat_eq}) describes a random
motion in $\mathbb{Q}_{p}$, more precisely, a Markov process with state space
$\mathbb{Q}_{p}$; see \cite{Zuniga-Textbook}, \cite{Kochubei}%
-\cite{Zuniga-LNM-2016}.

All operators $\boldsymbol{H}_{p}^{\text{free}}$ such that
(\ref{p_adic_Heat_eq}) is a $p$-adic heat equation are non-local. The paper
\cite{Zuniga-2026} demonstrates that by modeling microscopic space as a
totally disconnected topological space, the Schr\"{o}dinger equation governs
the dynamics of a system between measurements, while the region where the
wavefunction localizes during a measurement is fixed by the geometry of the
space (see Section \ref{Section_2}), without a postulate about observers.
Non-locality is built into the Hamiltonian rather than introduced ad hoc.
Additionally, the $p$-adic Dirac equation, which predicts the existence of
particle-antiparticle pairs and charge conjugation, admits localized
wavefunctions in a way that contradicts Einstein causality \cite{Zuniga-PhA};
this is a further indication that the framework is genuinely non-local. We
therefore argue that the mathematical non-locality arising from the use of
non-local operators is consistent with `phenomenological non-locality': the
existence of correlations between spatially separated particles that are
stronger than what any classical local theory permits, implying that measuring
a property of one particle can instantaneously influence the state of another
regardless of the distance between them.

Consequently, $\boldsymbol{H}$ is non-local. The standard QM on $L^{2}%
(\mathbb{R})$ is recovered by taking $\boldsymbol{H}_{p}=\boldsymbol{0}$,
while $p$-adic QM is recovered by taking $\boldsymbol{H}_{\infty
}=\boldsymbol{0}$.

\textbf{Realism.} In this paper, realism means the existence of a physical
reality independent of any observer. In the framework of QM on $L^{2}%
(\mathbb{R})\otimes L^{2}(\mathbb{Q}_{p})$, the observer-independent physical
entities are: (i) the definite readings (records) produced by the apparatuses,
and (ii) the probability measures $\left\vert \Psi\right\vert ^{2}dx_{\infty
}dx_{p}$ defined by the wavefunctions, which evolve deterministically and do
not depend on any observer. The wavefunction itself is a mathematical tool to
compute these measures; see Subsection \ref{Section_Further}. We do not claim
realism in the stronger sense of value definiteness, i.e., that all
observables (position, momentum, spin) possess definite pre-existing values
before any measurement, which is the sense used in the theorems of Bell and of
Kochen and Specker \cite{Bell64}, \cite{KochenSpecker}. In our framework, a
reading becomes definite when the wavefunction localizes during the
measurement interaction. Thus QM on $L^{2}(\mathbb{R})\otimes L^{2}%
(\mathbb{Q}_{p})$ is a non-local theory in which observed events are absolute:
they do not depend on any observer.

\subsection{Does the $p$-adic Schr\"{o}dinger equation describe a physical
system?}

QM\ on $\mathbb{C}^{N}$ has been used intensively in the construction of many
models of quantum systems. In \cite{Zuniga-2026}, \cite{Zuniga-QM-2}, the
author showed that QM\ on $\mathbb{C}^{N}$ can be recast as QM\ on
$L^{2}\left(  \mathbb{Z}_{p}\right)  $, where $\mathbb{Z}_{p}$ is the unit
ball in $\mathbb{Q}_{p}$. There exists a Hilbert space $\chi_{N}$ $\left(
\mathbb{Z}_{p}\right)  $ of dimension $N$ and embedding\ $\mathbb{C}^{N}%
\simeq\mathcal{\chi}_{N}(\mathbb{Z}_{p})\hookrightarrow L^{2}\left(
\mathbb{Z}_{p}\right)  $ such that for any Hermitian matrix $\left[
H_{i,j}\right]  _{N\times N}$ (a Hamiltonian in $\mathbb{C}^{N}$), there
exists a self-adjoint operator $\boldsymbol{H}_{p}:L^{2}\left(  \mathbb{Z}%
_{p}\right)  \rightarrow L^{2}\left(  \mathbb{Z}_{p}\right)  $ such that the
restriction operator $\boldsymbol{H}_{p}:\chi_{N}$ $\left(  \mathbb{Z}%
_{p}\right)  \rightarrow\chi_{N}$ $\left(  \mathbb{Z}_{p}\right)  $ is
well-defined, and the matrix corresponding is $\left[  H_{i,j}\right]
_{N\times N}$. We review this result in Appendix B. The meaning of this result
is that any discrete Schr\"{o}dinger equation on $\mathbb{C}^{N}$,
\[
\partial_{t}\left[  \psi_{i}\left(  t\right)  \right]  =\left[  H_{i,j}%
\right]  _{N\times N}\left[  \psi_{j}\left(  t\right)  \right]
\]
admits a continuous version of the form%
\[
\partial_{t}\,\psi_{p}(t,x_{p})\;=\;\boldsymbol{H}_{p}\,\psi_{p}%
(t,x_{p}),\qquad(t,x_{p})\in\mathbb{R}\times\mathbb{Z}_{p}.
\]

\section{\label{Section_2}The Collapse Mechanism}

The most important departure of \cite{Zuniga-2026} from standard QM is its
treatment of the wavefunction collapse. In standard QM, collapse is an
additional, non-dynamical postulate, and the questions of what counts as a
measurement and what counts as an observer are left open. In
\cite{Zuniga-2026}, a measurement is an interaction between a quantum system
and an apparatus, mediated by a continuous map from the microscopic space to
the macroscopic one, and the collapse is a localization of the wavefunction in
space, described by the standard projection rule, on a region determined by
the geometry of $\mathbb{R}\times\mathbb{Q}_{p}$. We now review and extend the
collapse mechanism introduced in \cite{Zuniga-2026}. In our framework,
$\mathbb{R}$ is a model for the physical macroscopic space. There are two
models for the microscopic space: $\mathbb{Q}_{p}$ and $\mathbb{R}%
\times\mathbb{Q}_{p}$. In \cite{Zuniga-2026}, only the first case was considered.

\subsection{\label{Section_Meas_I}The measurement problem I}

We consider a system consisting of two parts: a classical apparatus (a
macroscopic system) and a quantum object (a microscopic system). In this
section, we use $\mathbb{Q}_{p}$ as a model for the microscopic space. The
process of measurement involves the interaction of these two parts; as a
result the apparatus passes from its initial state into some other. From this
change of state one draws conclusions concerning the state of the quantum object.

Let $g$ be the pointer of the apparatus $\mathcal{A}$, let $\widehat
{\boldsymbol{g}}$ be the self-adjoint operator corresponding to the observable
$g$, which we suppose to have a discrete and non-degenerate spectrum. Let
$g_{n}\in\mathbb{R}$, $\theta_{n}\left(  x_{\infty}\right)  \in L^{2}%
(\mathbb{R})$, $n=1,2,\ldots$, be, respectively, the eigenvalues and
eigenfunctions of $\widehat{\boldsymbol{g}}$, where $L^{2}(\mathbb{R})$\ is
the state space of $\mathcal{A}$, and $\left\{  \theta_{n}\right\}  _{n}$ is
an orthonormal basis\ for $L^{2}(\mathbb{R})$. The wavefunctions of the
apparatus
\[
\Psi_{\mathcal{A}}\left(  t,x_{\infty}\right)  =%
{\displaystyle\sum\limits_{n=1}^{\infty}}
a_{n}\left(  t\right)  \theta_{n}\left(  x_{\infty}\right)
\]
are functions on the space-time $\left(  t,x_{\infty}\right)  \in
\mathbb{R}\times\mathbb{R}.$

Let $\psi_{n}$, $n=1,2,\ldots$ , be the eigenvectors of the operator
$\widehat{\boldsymbol{o}}$ corresponding to the observable $o$ that is to be
determined by the apparatus. We assume that $\left\{  \psi_{n}\right\}  _{n}$
is an orthonormal basis for $L^{2}(\mathbb{Q}_{p})$ consisting of wavelets,
see Appendix C; this is the case, for instance, when $\widehat{\boldsymbol{o}%
}$ is the Vladimirov operator, or a function of it, and its eigenvectors are
chosen as in Appendix C. This assumption is used in (\ref{Table-B}) below.
Then, the wavefunctions of the quantum system $\mathcal{S}$ have the form%
\[
\Psi_{\mathcal{S}}\left(  t,x_{p}\right)  =%
{\displaystyle\sum\limits_{n=1}^{\infty}}
b_{n}\left(  t\right)  \psi_{n}\left(  x_{p}\right)  .
\]
This means that they are functions on the space-time $\left(  t,x_{p}\right)
\in\mathbb{R}\times\mathbb{Q}_{p}$, and the state space of $\mathcal{S}$ is
$L^{2}(\mathbb{Q}_{p})$.

The system $\mathcal{S}+\mathcal{A}$ evolves under some unitary semigroup of
operators (a Schr\"{o}dinger equation) in the space $L^{2}(\mathbb{R})%
{\textstyle\bigotimes}
L^{2}(\mathbb{Q}_{p})$ starting from some initial state%
\[
\Psi_{\mathcal{A}}^{\left(  0\right)  }\left(  0,x_{\infty}\right)
{\textstyle\bigotimes}
\Psi_{\mathcal{S}}^{\left(  0\right)  }\left(  0,x_{p}\right)  \in
L^{2}(\mathbb{R})%
{\textstyle\bigotimes}
L^{2}(\mathbb{Q}_{p}).
\]
Now,
\[
L^{2}(\mathbb{R})\otimes L^{2}(\mathbb{Q}_{p})\simeq L^{2}(\mathbb{R\times
Q}_{p}),
\]
and $\left\{  \theta_{m}%
{\textstyle\bigotimes}
\psi_{n}\right\}  _{n,m}$ is an orthonormal basis for $L^{2}(\mathbb{R\times
Q}_{p})$; furthermore any element of $L^{2}(\mathbb{R\times Q}_{p})$ can be
uniquely represented as a series of the form%
\[%
{\displaystyle\sum\limits_{m=1}^{\infty}}
{\displaystyle\sum\limits_{n=1}^{\infty}}
c_{n,m}\psi_{n}\left(  x_{p}\right)  \theta_{m}\left(  x_{\infty}\right)  ,
\]
cf. \cite[Theorem II.10]{Reed-Simon-I}. From these considerations, the
wavefunction of $\mathcal{S}+\mathcal{A}$ has the form%
\begin{equation}
\Psi_{\mathcal{S}+\mathcal{A}}\left(  t,x_{\infty},x_{p}\right)  =%
{\displaystyle\sum\limits_{m=1}^{\infty}}
{\displaystyle\sum\limits_{n=1}^{\infty}}
c_{n,m}\left(  t\right)  \psi_{n}\left(  x_{p}\right)  \theta_{m}\left(
x_{\infty}\right)  . \label{Expansion}%
\end{equation}
The Born rule implies that\ $\ \left\vert \Psi_{\mathcal{S}+\mathcal{A}%
}\left(  t,x_{\infty},x_{p}\right)  \right\vert ^{2}$ is a probability
density, with respect to the measure $dx_{\infty}dx_{p}$, at the time $t$ in
the region $x_{\infty}\in I$, $x_{p}$ $\in B$, where $I\subset\mathbb{R}$ is
an interval and $B\subset\mathbb{Q}_{p}$ is a ball. It is important to mention
that, so far, $x_{\infty}$ and $x_{p}$ are independent degrees of freedom.

We use $\mathbb{Q}_{p}$ as a model for the microscopic space and $\mathbb{R}$
as a model for the macroscopic space. The interaction of the apparatus and the
quantum system requires a map
\begin{equation}
\mathcal{M}:\mathbb{Q}_{p}\rightarrow\mathbb{R}\text{,} \label{Monna_function}%
\end{equation}
which provides a distorted picture of quantum motion in macroscopic space. In
\cite{Zuniga-2026}, we use the Monna map, which is a simple and natural
choice. However, the author did not give a physical argument to support this
particular choice. So, it is natural to ask if the collapse mechanism's
apparent success may be an artifact of a particular map choice rather than a
deep structural consequence. To fix this problem, we assume that $\mathcal{M}$
is any continuous, surjective mapping. So our arguments will not depend on a
particular choice of $\mathcal{M}$. The condition that the preimage of a small
region in $\mathbb{R}$ should also be a small set in $\mathbb{Q}_{p}$
motivates our imposition of continuity for $\mathcal{M}$. Requiring
surjectivity ensures that every point in the model for macroscopic space is
represented. If this is not the case, we replace $\mathbb{R}$ by
$\mathcal{M}\left(  \mathbb{Q}_{p}\right)  $ endowed with the subspace topology.

Now, we take $I=\mathcal{M}(B)$, where $B$ is a ball (i.e., open compact
subset of $\mathbb{Q}_{p}$), so, $I$ is a compact subset of $\mathbb{R}$. Then%
\begin{equation}
P_{\text{int}}(B,t)=\frac{1}{A\left(  \mathcal{M}\right)  }\int\limits_{B}%
\left\vert \Psi_{\mathcal{S}+\mathcal{A}}\left(  t,\mathcal{M}(x_{p}%
),x_{p}\right)  \right\vert ^{2}dx_{p} \label{P_int}%
\end{equation}
is the probability of interaction of the apparatus and the quantum object at
the time $t$ in the region $B$, where%
\[
A\left(  \mathcal{M}\right)  =\int\limits_{\mathbb{Q}_{p}}\left\vert
\Psi_{\mathcal{S}+\mathcal{A}}\left(  t,\mathcal{M}(x_{p}),x_{p}\right)
\right\vert ^{2}dx_{p}.
\]
Formula (\ref{P_int}) involves the amplitude $\varphi_{t}(x_{p}):=\Psi
_{\mathcal{S}+\mathcal{A}}\left(  t,\mathcal{M}(x_{p}),x_{p}\right)  $ of the
state along the graph of $\mathcal{M}$, which is a function in $L^{2}%
(\mathbb{Q}_{p})$; by Remark \ref{Remark_graph} below, it is the restriction
to the graph of the element $\left(  1_{\mathcal{M}(B)}\otimes1_{B}\right)
\Psi_{\mathcal{S}+\mathcal{A}}$ of $L^{2}(\mathbb{R})\otimes L^{2}%
(\mathbb{Q}_{p})$. In our framework, the map $\mathcal{M}$ specifies the
interpretation of the measurement: at the end of the interaction, the state of
the quantum object probed by the apparatus is $\varphi_{T}$, and (\ref{P_int})
and the probability of the reading given below are the Born probabilities, in
$L^{2}(\mathbb{Q}_{p})$, of the projections $1_{B}$ and $1_{\mathcal{M}%
^{-1}(L)}$.

\begin{remark}
The specific ball $B$ scanned by the apparatus is not chosen a priori; it is
determined during the measurement process. In our model of quantum
measurement, the apparatus scans a bounded region $J$ (an open interval) in
$\mathbb{R}$. Then $\mathcal{M}^{-1}(J)=A$ is an open subset. We take $B$ a
ball contained in $A$, and set $\mathcal{M}(B)=I\subset J$. Notice that we do
not require the radius of $B$ to be sufficiently small; we just need any ball
contained in $A$. This observation assures that there is no conflict with the
existence of the Planck length or with any uncertainty principle limiting the
radius of $B$. The ball $B$ is selected according to the probabilities
(\ref{P_int}); in this sense, the Born rule is retained in our framework.
\end{remark}

In the next step, we compute an explicit expression for
\begin{equation}
1_{B}\left(  x_{p}\right)  \Psi_{\mathcal{S}+\mathcal{A}}\left(
t,\mathcal{M}(x_{p}),x_{p}\right)  =%
{\displaystyle\sum\limits_{m=1}^{\infty}}
{\displaystyle\sum\limits_{n=1}^{\infty}}
c_{n,m}\left(  t\right)  \left\{  1_{B}\left(  x_{p}\right)  \psi_{n}\left(
x_{p}\right)  \right\}  \theta_{m}\left(  \mathcal{M}(x_{p})\right)  ,
\label{Expansion-B}%
\end{equation}
where $1_{B}\left(  x_{p}\right)  $ is the characteristic function of ball
$B$. In \cite{Zuniga-2026}, \ we use an orthonormal basis consisting of
eigenfunctions of the Vladimirov operator; the calculations are reviewed in
Appendix C. Here, we summarize these calculations using the basis $\left\{
\psi_{n}\left(  x_{p}\right)  \right\}  _{n}$. We pick the $B=p^{l}%
a+p^{l}\mathbb{Z}_{p}$, and assume that $\psi_{n}\left(  x_{p}\right)  $ is
supported in $bp^{-r}+p^{-r}\mathbb{Z}_{p}$, where $l$, $a$ \ are fixed
parameters, while $b$, $r$ depend on $n$. It is more convenient using the
notation $1_{B}\left(  x_{p}\right)  =$ $\Omega\left(  p^{l}\left\vert
x_{p}-p^{l}a\right\vert _{p}\right)  $. The key calculation shows that
\begin{equation}
\Omega\left(  p^{l}\left\vert x_{p}-p^{l}a\right\vert _{p}\right)  \psi
_{n}\left(  x_{p}\right)  =\left\{
\begin{array}
[c]{ll}%
\psi_{n}\left(  x_{p}\right)  & \text{if }n\in\mathcal{N}_{1}\\
& \\
c_{n}^{\prime}\Omega\left(  p^{l}\left\vert x_{p}-p^{l}a\right\vert
_{p}\right)  & \text{if }n\in\mathcal{N}_{2}\\
& \\
0 & \text{if }n\in\mathcal{N}_{3},
\end{array}
\right.  \label{Table-B}%
\end{equation}
where $c_{n}^{\prime}$ is a complex number depending on $n$, and the sets
$\mathcal{N}_{1}$, $\mathcal{N}_{2}$, $\mathcal{N}_{3}$ are disjoint. This
calculation uses the fact that in the $p$-adic topology, two balls are
disjoint or one is contained in the other; this assertion is false in
$\mathbb{R}$. These facts, and the explicit form of (\ref{Table-B}) for the
basis of wavelets, are standard results of $p$-adic analysis; see
\cite[Theorems 9.4.5 and 8.9.3]{Alberio et al}, \cite[Theorem 3.3]{KKZuniga},
\cite[Table 4.4]{Zuniga-PhA}, and Appendix C. The first line in (\ref{Table-B}%
) covers the case \textrm{supp}$\psi_{n}\left(  x_{p}\right)  \subseteq
$\textrm{supp}$\Omega\left(  p^{l}\left\vert x_{p}-p^{l}a\right\vert
_{p}\right)  $, while the second one corresponds to \textrm{supp}%
$\Omega\left(  p^{l}\left\vert x_{p}-p^{l}a\right\vert _{p}\right)  \subset
$\textrm{supp}$\psi_{n}\left(  x_{p}\right)  $, and the last line corresponds
to
\[
\mathrm{supp}\psi_{n}\left(  x_{p}\right)  \cap\mathrm{supp}\Omega\left(
p^{l}\left\vert x_{p}-p^{l}a\right\vert _{p}\right)  =\emptyset.
\]
Notice that the first line in (\ref{Table-B}) says $\Omega\left(
p^{l}\left\vert x_{p}-p^{l}a\right\vert _{p}\right)  \psi_{n}\left(
x_{p}\right)  =\psi_{n}\left(  x_{p}\right)  $ for $n\in\mathcal{N}_{1}$.
Then, using (\ref{Expansion-B}),
\begin{gather}
\Omega\left(  p^{l}\left\vert x_{p}-p^{l}a\right\vert _{p}\right)
\Psi_{\mathcal{S}+\mathcal{A}}\left(  t,\mathcal{M}(x_{p}),x_{p}\right)
=\label{Wavefunction}\\
\Omega\left(  p^{l}\left\vert x_{p}-p^{l}a\right\vert _{p}\right)  \left\{
{\displaystyle\sum\limits_{m=1}^{\infty}}
\text{ }%
{\displaystyle\sum\limits_{n\in\mathcal{N}_{1}}}
c_{n,m}\left(  t\right)  \psi_{n}\left(  x_{p}\right)  \theta_{m}\left(
\mathcal{M}\left(  x_{p}\right)  \right)  \right. \nonumber\\
\left.  +%
{\displaystyle\sum\limits_{m=1}^{\infty}}
\text{ }%
{\displaystyle\sum\limits_{n\in\mathcal{N}_{2}}}
c_{n}^{\prime}c_{n,m}\left(  t\right)  \theta_{m}\left(  \mathcal{M}\left(
x_{p}\right)  \right)  \right\}  =\nonumber\\
\Omega\left(  p^{l}\left\vert x_{p}-p^{l}a\right\vert _{p}\right)
{\displaystyle\sum\limits_{m=1}^{\infty}}
A_{m}(t,x_{p})\theta_{m}\left(  \mathcal{M}\left(  x_{p}\right)  \right)
,\text{ }x_{p}\in B.\nonumber
\end{gather}
where $A_{m}(t,x_{p}):=\sum_{n\in\mathcal{N}_{1}}c_{n,m}(t)\psi_{n}%
(x_{p})+\sum_{n\in\mathcal{N}_{2}}c_{n}^{\prime}c_{n,m}(t)$.

We assume that the interaction between the apparatus and the quantum particle
starts at $t=0$ and ends at $t=T$. Then, using (\ref{Wavefunction}), the
wavefunction of $\mathcal{S}+\mathcal{A}$ on the region of interaction at the
time $t=T$ is%
\[
\Psi_{\mathcal{S}+\mathcal{A}}\left(  T,\mathcal{M}\left(  x_{p}\right)
,x_{p}\right)  =\sum\limits_{m=1}^{\infty}A_{m}(T,x_{p})\theta_{m}\left(
\mathcal{M}\left(  x_{p}\right)  \right)  \text{, for }x_{p}\in B,
\]
which is a function supported on a compact subset. The wavefunction collapses
in space. The probability that the reading of the apparatus be in an interval
$L\subset I=\mathcal{M}\left(  B\right)  $ is%
\[
\frac{\int\limits_{\mathcal{M}^{-1}\left(  L\right)  \cap B}\left\vert
\Psi_{\mathcal{S}+\mathcal{A}}\left(  T,\mathcal{M}\left(  x_{p}\right)
,x_{p}\right)  \right\vert ^{2}dx_{p}}{\int\limits_{B}\left\vert
\Psi_{\mathcal{S}+\mathcal{A}}\left(  T,\mathcal{M}\left(  x_{p}\right)
,x_{p}\right)  \right\vert ^{2}dx_{p}}.
\]

\begin{remark}
\label{Remark_graph}We now discuss the interpretation of (\ref{Wavefunction}).
Its last line says that, on the region of interaction, the wavefunction is
localized in position: it is supported in the compact set $\left\{
(\mathcal{M}(x_{p}),x_{p});x_{p}\in B\right\}  $. Formula (\ref{Wavefunction})
involves the values of $\Psi_{\mathcal{S}+\mathcal{A}}$ on the graph of
$\mathcal{M}$, which is a set of measure zero in $\mathbb{R}\times
\mathbb{Q}_{p}$. To relate it with states in $L^{2}(\mathbb{R})\otimes
L^{2}(\mathbb{Q}_{p})$, we apply to (\ref{Expansion}) the orthogonal
projection $1_{\mathcal{M}(B)}\otimes1_{B}$, given by multiplication by
$1_{\mathcal{M}(B)}(x_{\infty})1_{B}(x_{p})$. By (\ref{Table-B}),%
\begin{multline}
\left(  1_{\mathcal{M}(B)}\otimes1_{B}\right)  \Psi_{\mathcal{S}+\mathcal{A}%
}\left(  t,x_{\infty},x_{p}\right)  =\sum_{m=1}^{\infty}\sum_{n\in
\mathcal{N}_{1}}c_{n,m}(t)\,\psi_{n}(x_{p})\,1_{\mathcal{M}(B)}(x_{\infty
})\theta_{m}(x_{\infty})\\
+\sum_{m=1}^{\infty}\sum_{n\in\mathcal{N}_{2}}c_{n}^{\prime}c_{n,m}%
(t)\,1_{B}(x_{p})\,1_{\mathcal{M}(B)}(x_{\infty})\theta_{m}(x_{\infty}).
\label{Wavefunction2}%
\end{multline}
This is an element of $L^{2}(\mathbb{R})\otimes L^{2}(\mathbb{Q}_{p})$
supported in the compact set $\mathcal{M}(B)\times B$, and its restriction to
the graph of $\mathcal{M}$ over $B$ is (\ref{Wavefunction}). Thus,
(\ref{Wavefunction}) is the restriction to the graph of $\mathcal{M}$ of a
genuine element of $L^{2}(\mathbb{R})\otimes L^{2}(\mathbb{Q}_{p})$.
\end{remark}

\textbf{The update rule.} At the end of the interaction, the state of
$\mathcal{S}+\mathcal{A}$ is replaced by%
\begin{equation}
\Psi_{\mathcal{S}+\mathcal{A}}^{B}\left(  x_{p}\right)  =\frac{1_{B}\left(
x_{p}\right)  \Psi_{\mathcal{S}+\mathcal{A}}\left(  T,\mathcal{M}\left(
x_{p}\right)  ,x_{p}\right)  }{\left(  \int_{B}\left\vert \Psi_{\mathcal{S}%
+\mathcal{A}}\left(  T,\mathcal{M}\left(  y\right)  ,y\right)  \right\vert
^{2}dy\right)  ^{1/2}}, \label{Update_rule}%
\end{equation}
where the ball $B$ is selected with probability $P_{\text{int}}(B,T)$. This is
the standard projection rule of the Dirac--von Neumann formalism, applied to
the orthogonal projection $1_{B}$ on $L^{2}(\mathbb{Q}_{p})$ and to the state
$\varphi_{T}$. Since the subsequent evolution of $\mathcal{S}+\mathcal{A}$
takes place in $L^{2}(\mathbb{R})\otimes L^{2}(\mathbb{Q}_{p})$, we take as
initial condition for this evolution the state%
\begin{equation}
\widetilde{\Psi}_{\mathcal{S}+\mathcal{A}}^{B}=\frac{\left(  1_{\mathcal{M}%
(B)}\otimes1_{B}\right)  \Psi_{\mathcal{S}+\mathcal{A}}(T)}{\left\Vert \left(
1_{\mathcal{M}(B)}\otimes1_{B}\right)  \Psi_{\mathcal{S}+\mathcal{A}%
}(T)\right\Vert }\in L^{2}(\mathbb{R})\otimes L^{2}(\mathbb{Q}_{p}),
\label{Lift}%
\end{equation}
i.e., the normalization of (\ref{Wavefunction2}) at $t=T$; by Remark
\ref{Remark_graph}, its restriction to the graph of $\mathcal{M}$ is
proportional to (\ref{Update_rule}). The framework does not introduce a new
update rule; it specifies when the rule applies and which projection is
involved: the projection takes place when a quantum system interacts with an
apparatus through the map $\mathcal{M}$, and it localizes the state on the
graph of $\mathcal{M}$ over $B$, where $B$ is a ball of $\mathbb{Q}_{p}$. No
observer is involved. Between measurements, the state evolves according to the
Schr\"{o}dinger equation. When several measurements are performed
consecutively, the update rule, in the form (\ref{Lift}), provides the initial
condition for the Schr\"{o}dinger equation that describes the evolution up to
the next measurement. The collapse mechanism itself, i.e., the localization
(\ref{Wavefunction}) on the region of interaction, does not depend on the
update rule: every measurement produces a definite reading of the pointer. The
alternative reading, in which the localization is only a projection while the
global state continues its unitary evolution, is discussed in Subsection
\ref{Section_Further} and in Appendix F.

\textbf{Apparatuses and quantum systems.} A system acts as an apparatus in a
measurement when it interacts with the quantum system $\mathcal{S}$ through a
map $\mathcal{M}$ from the microscopic space into the macroscopic one; this is
the situation in which the update rule (\ref{Update_rule}) applies. A system
that does not interact with $\mathcal{S}$ through such a map behaves as a
quantum system: its evolution is unitary. This criterion, and its consequences
for extended Wigner's Friend experiments, are discussed in Subsection
\ref{Section_App}.

\textbf{An example.} In Appendix D, we illustrate the mechanism with a toy
model: the measurement of the energy levels of a particle in a box. In
standard QM on $L^{2}(\mathbb{R})$, the pointer is left in a superposition of
its possible positions. In QM on $L^{2}(\mathbb{R}\times\mathbb{Q}_{p})$, when
the apparatus scans the interval $\left[  p^{-n},2p^{-n}\right]
=\mathcal{M}\left(  p^{n-1}+p^{n}\mathbb{Z}_{p}\right)  $, the pointer
wavefunction is localized there, up to exponentially small terms, and the
reading corresponds to the energy level $p^{n}$.

\subsection{\label{Section_Meas_II}The measurement problem II}

We now consider a system consisting of two parts: a classical apparatus (a
macroscopic system) $\mathcal{S}_{2}$ and a quantum system $\mathcal{S}_{1}$.
In this section, we use $\mathbb{R}\times\mathbb{Q}_{p}$ as a model for the
microscopic space. The state space for $\mathcal{S}_{1}$ is now $L^{2}%
(\mathbb{R})%
{\textstyle\bigotimes}
L^{2}(\mathbb{Q}_{p})\simeq L^{2}(\mathbb{R\times Q}_{p})$, and $\left\{
\theta_{m}%
{\textstyle\bigotimes}
\psi_{n}\right\}  _{n,m}$ is an orthonormal basis for $L^{2}(\mathbb{R\times
Q}_{p})$. Then, the wavefunction of system $\mathcal{S}_{1}$ has the form%
\[
\Psi_{\mathcal{S}_{1}}\left(  t,x_{\infty},x_{p}\right)  =%
{\displaystyle\sum\limits_{m=1}^{\infty}}
{\displaystyle\sum\limits_{n=1}^{\infty}}
a_{n,m}\left(  t\right)  \psi_{n}\left(  x_{p}\right)  \theta_{m}\left(
x_{\infty}\right)  .
\]
We assume that the state space for system $\mathcal{S}_{2}$ is $L^{2}%
(\mathbb{R})$. The wavefunctions of the apparatus have the form
\[
\Psi_{\mathcal{S}_{2}}\left(  t,y_{\infty}\right)  =%
{\displaystyle\sum\limits_{r=1}^{\infty}}
b_{r}\left(  t\right)  \omega_{r}\left(  y_{\infty}\right)  ,
\]
where $\left\{  \omega_{r}\right\}  _{r}$ is an orthonormal basis for
$L^{2}(\mathbb{R})$.

Starting at an initial state from $L^{2}(\mathbb{R\times Q}_{p})%
{\textstyle\bigotimes}
L^{2}(\mathbb{R})$, the system $\mathcal{S}_{1}+\mathcal{S}_{2}$ evolves in
this space with a wavefunction of the form%
\begin{equation}
\Psi_{\mathcal{S}_{1}+\mathcal{S}_{2}}\left(  t,x_{\infty},x_{p},y_{\infty
}\right)  =%
{\displaystyle\sum\limits_{r=1}^{\infty}}
{\displaystyle\sum\limits_{m=1}^{\infty}}
{\displaystyle\sum\limits_{n=1}^{\infty}}
d_{n,m,r}\left(  t\right)  \psi_{n}\left(  x_{p}\right)  \omega_{r}\left(
y_{\infty}\right)  \theta_{m}\left(  x_{\infty}\right)  . \label{Expansion_2}%
\end{equation}
The Born rule implies that\ $\left\vert \Psi_{\mathcal{S}_{1}+\mathcal{S}_{2}%
}\left(  t,x_{\infty},x_{p},y_{\infty}\right)  \right\vert ^{2}$ is a
probability density, with respect to the measure $dx_{\infty}dx_{p}dy_{\infty
}$, at the time $t$ in the region $x_{\infty}\in I$, $x_{p}\in B$, and
$y_{\infty}\in J$. The interaction of the systems $\mathcal{S}_{1}$ and
$\mathcal{S}_{2}$, which have different configuration spaces ($\mathbb{R\times
Q}_{p}$, respectively $\mathbb{R}$) requires a continuous map
\[
\mathcal{M}_{\ast}:\mathbb{R\times Q}_{p}\rightarrow\mathbb{R}\text{,}%
\]
which is the analog of (\ref{Monna_function}). We take a small open interval
$J\subset\mathbb{R}$, and a set $I\times B$ contained in the open subset
$\mathcal{M}_{\ast}^{-1}(J)$, where $I$ is a compact subset of $\mathbb{R}$
and $B$ is a ball in $\mathbb{Q}_{p}$; so $\mathcal{M}_{\ast}\left(  I\times
B\right)  $ is a compact subset contained in $J$. As in Subsection
\ref{Section_Meas_I}, the interaction is described by the amplitude of the
state along the graph of $\mathcal{M}_{\ast}$,%
\[
\varphi_{t}^{\ast}(x_{\infty},x_{p}):=\Psi_{\mathcal{S}_{1}+\mathcal{S}_{2}%
}\left(  t,x_{\infty},x_{p},\mathcal{M}_{\ast}\left(  x_{\infty},x_{p}\right)
\right)  ,
\]
which is a function in $L^{2}(\mathbb{R})\otimes L^{2}(\mathbb{Q}_{p})$. The
probability of interaction at time $t$ on the region $I\times B$ is%
\[
P_{\text{int}}(I\times B,t)=\frac{1}{C\left(  \mathcal{M}_{\ast}\right)  }%
\int\limits_{I}\int\limits_{B}\left\vert \varphi_{t}^{\ast}(x_{\infty}%
,x_{p})\right\vert ^{2}dx_{p}dx_{\infty}=\frac{\left\Vert \left(  1_{I}%
\otimes1_{B}\right)  \varphi_{t}^{\ast}\right\Vert ^{2}}{\left\Vert
\varphi_{t}^{\ast}\right\Vert ^{2}},
\]
where $C\left(  \mathcal{M}_{\ast}\right)  =\left\Vert \varphi_{t}^{\ast
}\right\Vert ^{2}$, and $1_{I}\otimes1_{B}$ denotes the orthogonal projection
on $L^{2}(\mathbb{R})\otimes L^{2}(\mathbb{Q}_{p})$ given by multiplication by
$1_{I}(x_{\infty})1_{B}(x_{p})$. In this formula, $x_{\infty}$ and $x_{p}$ are
independent degrees of freedom, and the sets $I$ and $B$ are unrelated. It is
crucial to note that we cannot establish a space localization similar to
(\ref{Wavefunction}) unless we introduce an extra hypothesis that relates $I$
and $B$. This hypothesis is precisely the existence of the map $\mathcal{M}$,
see (\ref{Monna_function}), which gives $I=\mathcal{M}(B)$ and ties
$x_{\infty}$ to $x_{p}$ through $x_{\infty}=\mathcal{M}(x_{p})$. We set%
\[
G(x_{p}):=\left(  \mathcal{M}(x_{p}),x_{p},\mathcal{M}_{\ast}\left(
\mathcal{M}(x_{p}),x_{p}\right)  \right)  ,\qquad\varphi_{t}^{\mathcal{W}%
}(x_{p}):=\Psi_{\mathcal{S}_{1}+\mathcal{S}_{2}}\left(  t,G(x_{p})\right)
=\varphi_{t}^{\ast}\left(  \mathcal{M}(x_{p}),x_{p}\right)  ,
\]
so that $\varphi_{t}^{\mathcal{W}}\in L^{2}(\mathbb{Q}_{p})$ is the amplitude
of the state along the set $\left\{  G(x_{p});x_{p}\in\mathbb{Q}_{p}\right\}
$. With the notation introduced in Subsection \ref{Section_Meas_I},%
\[
P_{\text{int}}(B,t)=\frac{1}{C\left(  \mathcal{M},\mathcal{M}_{\ast}\right)
}\int\limits_{B}\left\vert \varphi_{t}^{\mathcal{W}}(x_{p})\right\vert
^{2}dx_{p}=\frac{\left\Vert 1_{B}\varphi_{t}^{\mathcal{W}}\right\Vert ^{2}%
}{\left\Vert \varphi_{t}^{\mathcal{W}}\right\Vert ^{2}},
\]
where $C\left(  \mathcal{M},\mathcal{M}_{\ast}\right)  =\left\Vert \varphi
_{t}^{\mathcal{W}}\right\Vert ^{2}$ is the same integral taken over
$\mathbb{Q}_{p}$. As in Subsection \ref{Section_Meas_I}, the maps
$\mathcal{M}$ and $\mathcal{M}_{\ast}$ specify the interpretation of the
measurement: at the end of the interaction, the state of the system probed by
$\mathcal{S}_{2}$ is $\varphi_{T}^{\mathcal{W}}$, and $P_{\text{int}}(B,T)$ is
the Born probability, in $L^{2}(\mathbb{Q}_{p})$, of the projection $1_{B}$.
The function $\varphi_{t}^{\mathcal{W}}$ admits a localization similar to
(\ref{Wavefunction}). The probability that the reading of $\mathcal{S}_{2}$ be
in an interval $L\subset\mathcal{M}_{\ast}\left(  \mathcal{M}(B)\times
B\right)  $ is%
\[
\frac{\int\limits_{g^{-1}(L)\cap B}\left\vert \varphi_{T}^{\mathcal{W}}%
(x_{p})\right\vert ^{2}dx_{p}}{\int\limits_{B}\left\vert \varphi
_{T}^{\mathcal{W}}(x_{p})\right\vert ^{2}dx_{p}},\qquad g(x_{p}):=\mathcal{M}%
_{\ast}\left(  \mathcal{M}(x_{p}),x_{p}\right)  ,
\]
and the update rule (\ref{Update_rule}) applies in the form%
\begin{equation}
\Psi_{\mathcal{S}_{1}+\mathcal{S}_{2}}^{B}(x_{p})=\frac{1_{B}(x_{p}%
)\varphi_{T}^{\mathcal{W}}(x_{p})}{\left\Vert 1_{B}\varphi_{T}^{\mathcal{W}%
}\right\Vert },\label{Update_rule_W}%
\end{equation}
i.e., with the graph of $\mathcal{M}$ over $B$ replaced by the set $\left\{
G(x_{p});x_{p}\in B\right\}  $. As in (\ref{Lift}), the initial condition for
the subsequent evolution in $L^{2}(\mathbb{R}\times\mathbb{Q}_{p})\otimes
L^{2}(\mathbb{R})$ is the state%
\begin{equation}
\widetilde{\Psi}_{\mathcal{S}_{1}+\mathcal{S}_{2}}^{B}=\frac{\left(
1_{\mathcal{M}(B)}\otimes1_{B}\otimes1_{g(B)}\right)  \Psi_{\mathcal{S}%
_{1}+\mathcal{S}_{2}}(T)}{\left\Vert \left(  1_{\mathcal{M}(B)}\otimes
1_{B}\otimes1_{g(B)}\right)  \Psi_{\mathcal{S}_{1}+\mathcal{S}_{2}%
}(T)\right\Vert },\label{Lift_W}%
\end{equation}
where $1_{\mathcal{M}(B)}\otimes1_{B}\otimes1_{g(B)}$ denotes
multiplication by $1_{\mathcal{M}(B)}(x_{\infty})1_{B}(x_{p})1_{g(B)}%
(y_{\infty})$.

\begin{remark}
\label{Nota1}The above argument is valid if we replace the state space for
system $\mathcal{S}_{2}$ by $L^{2}(\mathbb{R}^{N+1})$. The meaning of this
choice is that the apparatus is entangled with $N$ macroscopic systems. These
macroscopic systems interact with the quantum system only through
$\mathcal{M}_{\ast}$, and the criterion of Subsection \ref{Section_App}
applies to the apparatus as a whole; entanglement among macroscopic degrees of
freedom does not prevent the localization. For Wigner to observe interference
it would be necessary to reverse the localization, which the update rule
(\ref{Update_rule}) excludes.
\end{remark}

\subsection{\label{Section_Further}Further comments}

\subsubsection{Comparison with collapse models and Bohmian mechanics}

The Ghirardi--Rimini--Weber (GRW) theory posits that the collapse of the
wavefunction occurs in position space, meaning that the wavefunction localizes
to a specific region in space, see, e.g., \cite{Norsen}-\cite{Bell-q-Jumps};
the continuous spontaneous localization (CSL) models \cite{Pearle-CSL}-\cite{GPR90} replace the discrete jumps of GRW by a continuous stochastic
process; see \cite{Bassi} for a review. In these theories, the wavefunctions
are physical entities that undergo a spontaneous, random collapse in nature.
This requires modifying the Schr\"{o}dinger equation by nonlinear and
stochastic terms, and introducing new physical constants (a collapse rate and
a localization length).

The localization process proposed in \cite{Zuniga-2026} resembles that of the
GRW theory, because the wavefunction of the apparatus localizes during the
measurement. However, in \cite{Zuniga-2026}, the wavefunctions are not
physical entities; the physical entities are the probability measures that
they define and the readings of the apparatuses (see the paragraph on realism
in Section \ref{Section_1}). It is not necessary to introduce new physical
constants, nor to replace the Schr\"{o}dinger equation with another equation
involving nonlinear and stochastic terms. The localization is described by the
standard projection rule; it occurs only during the interaction of a quantum
system with an apparatus through the map $\mathcal{M}$, and the region of
localization, the graph of $\mathcal{M}$ over the ball $B$, is determined by
the geometry of the macroscopic and microscopic spaces. The selection of the
ball $B$ is governed by the Born rule; there is no stochastic dynamical process.

The difference with the GRW theory can be stated more precisely. In QM on
$L^{2}(\mathbb{R}\times\mathbb{Q}_{p})$, with $\mathbb{R}$ a model of the
macroscopic space and $\mathbb{Q}_{p}$ a model of the microscopic space, the
map $\mathcal{M}:\mathbb{Q}_{p}\rightarrow\mathbb{R}$ appears as a parameter.
This parameter does not change the Dirac--von Neumann formulation: the Hilbert
space, the observables, the Schr\"{o}dinger evolution, and the Born rule are
the standard ones. It changes the interpretation of quantum measurements: at
the end of a measurement interaction, the state of the probed quantum object
is the amplitude of the state along the graph of $\mathcal{M}$, and the
localization takes place on this graph. This point is discussed in
\cite{Zuniga-2026} and the references therein. In contrast, the GRW and CSL
theories modify the dynamics itself.

Bohmian mechanics \cite{Bohm52} is the closest structural relative of our
framework: in both, non-locality is explicit, and definite outcomes are
obtained without observers. In Bohmian mechanics, definiteness comes from a
privileged configuration (the positions of the particles), guided by a
wavefunction that never collapses; the effective collapse is described by the
conditional wavefunction. In our framework, there is no additional variable:
definiteness comes from the localization of the state on a ball of
$\mathbb{Q}_{p}$. We notice that the Monna map is not injective, so the
$p$-adic coordinate $x_{p}$ carries information that is not visible in the
macroscopic coordinate; whether $x_{p}$ can play a role analogous to the
configuration variables of Bohmian mechanics is an open question. For a
discussion of localization in relativistic quantum field theory, see
\cite{Falcone-Conti}.

The following table summarizes the comparison.

\begin{center}
{\small
\begin{tabular}
[c]{|l|c|c|c|c|}\hline
Feature & Standard QM & GRW/CSL & Bohm & This framework\\\hline
Modifies Schr\"{o}dinger eq. & no & yes & no & no\\\hline
New constants & no & yes & no & no\\\hline
Collapse & postulate & stochastic & effective & postulate, via $\mathcal{M}%
$\\\hline
Additional variables & no & no & positions & no\\\hline
Role of the observer & unspecified & none & none & none\\\hline
\end{tabular}
}
\end{center}

\subsubsection{Localization as a projection}

Two operations are involved in a measurement. First, the state $\Psi
_{\mathcal{S}+\mathcal{A}}\in L^{2}(\mathbb{R})\otimes L^{2}(\mathbb{Q}_{p})$
is replaced by its amplitude $\varphi_{T}$ along the graph of $\mathcal{M}$,
an element of $L^{2}(\mathbb{Q}_{p})$; by Remark \ref{Remark_graph},
$\varphi_{T}$ is the restriction to the graph of an element of $L^{2}%
(\mathbb{R})\otimes L^{2}(\mathbb{Q}_{p})$. Second, the orthogonal projection
$1_{B}$ is applied to $\varphi_{T}$, according to the update rule
(\ref{Update_rule}). The first operation is a projection from the
higher-dimensional configuration space $\mathbb{R}\times\mathbb{Q}_{p}$ onto
the graph of $\mathcal{M}$; it is the source of the localization
(\ref{Wavefunction}), and hence of the definiteness of the reading, which, as
we noted in Subsection \ref{Section_Meas_I}, does not depend on the update rule.

One may ask whether the update rule should be regarded as a physical change of
state, as we do in this paper, or whether the global state should be regarded
as evolving unitarily on the full configuration space, the superposition never
being removed, while the localized state only describes the reading. We adopt
the first alternative. In the second alternative, an apparatus that later
interacts with the global state, such as Wigner in the Wigner's Friend
scenario of Section \ref{Section_4}, could in principle observe interference between the components corresponding
to the different readings of the Friend, and the original paradox would
reappear, unless one shows that Wigner's measurement, mediated by
$\mathcal{M}$ and $\mathcal{M}_{\ast}$ (Subsection \ref{Section_Meas_II}),
also removes the interference terms. This alternative is analyzed in Appendix
F: the ultrametric geometry excludes interference terms in Wigner's statistics
provided that the dynamics between the two measurements does not move
amplitude between the balls corresponding to the different readings of the
Friend; when it does, the record of the Friend's reading can be undone (see
Subsection \ref{Section_Persist}). The update rule (\ref{Update_rule}) is
precisely what guarantees the persistence of the Friend's record in all
cases.

\subsubsection{Non-locality and the no-communication theorem}

We want to point out an interesting connection between the work of K. K. Wan
\cite{Wan} and \cite{Zuniga-2026}. Both papers reject the idea that quantum
mechanics can be understood with a uniform spatial ontology across all scales.
Instead, each argues that short-distance physics must be fundamentally
non-local, while large-distance or macroscopic behavior recovers a more
classical, separable character. Beyond this shared intuition, however, the
papers diverge sharply in their motivations, mathematical architectures, and
scope. In \cite{Wan}, the algebraic, continuous-space framework is not in
conflict with special relativity per se; indeed, Haag--Kastler-type algebras
were developed precisely to be relativistically covariant. On the other hand,
in \cite{Zuniga-2026}, see also \cite{Zuniga-PhA}, because a totally
disconnected space has no continuous curves, the framework is explicitly
incompatible with special relativity at the microscopic level. The new theory
lacks Lorentz symmetry, and Einstein causality is violated: the $p$-adic Dirac
equation admits localized solutions with superluminal transition probabilities
\cite{Zuniga-PhA}. This fact does not contradict the so-called
no-communication theorem; such a result requires, as a primary hypothesis,
that $\mathbb{R}^{4}$ be a valid model for space-time at the Planck scale.
Thus, the no-communication theorem under the discreteness of the space is an
open problem. In Section \ref{Section_5} we show that, when no apparatus
localization occurs, the statistics are those of standard QM and are
no-signalling; whether no-signalling holds at the level of observable
statistics when apparatus localizations are involved, in view of the results
of \cite{Zuniga-PhA}, remains open.

\section{\label{Section_4}The Wigner's Friend Paradox in the Framework of
Space-discreteness Hypothesis}

\subsection{The Wigner's Friend Paradox}

The Wigner's Friend Paradox is a QM thought experiment that demonstrates that
two different observers can experience completely different, yet
mathematically valid, realities \cite{Wigner61}. The paradox expands upon
Schr\"{o}dinger's Cat to highlight the measurement problem. The paradox
unfolds in two steps, comparing the perspectives of two observers---Wigner and
His Friend:

\begin{itemize}
\item Inside the Lab: Wigner's friend is inside a sealed laboratory measuring
a quantum system (e.g., a photon in a superposition of states). From the
friend's perspective, the measurement is made, the wavefunction collapses, and
the photon acquires a definite state.

\item Outside the Lab: Wigner is waiting outside the sealed lab. Because the
lab is perfectly isolated, Wigner considers the entire lab---the measuring
device, the photon, and the friend---to be a single, massive quantum system.
To Wigner, all these elements remain in a superposition of all possible
outcomes until he opens the door and observes the lab's interior.
\end{itemize}

The problem arises when we ask: When did the wavefunction actually collapse?
According to the friend, it collapsed the moment they observed the photon. But
according to Wigner, it did not collapse until he looked inside the lab. Both
observers are applying the mathematically correct laws of quantum mechanics,
yet they arrive at mutually exclusive descriptions of reality at the same
point in time. For decades, this paradox was simply a philosophical puzzle
regarding the nature of "measurement" and whether consciousness is required to
collapse a wavefunction. However, modern extensions of the paradox have been
tested experimentally \cite{Proietti}, \cite{Bong et al}. These extended
scenarios force us to abandon at least one of three assumptions: the
Absoluteness of Observed Events, Locality, and No-Superdeterminism (freedom of
choice) \cite{Bong et al}.

\subsection{Solution of the Wigner's Friend Paradox}

In this Section, we apply the collapse mechanism for QM on $L^{2}%
(\mathbb{R\times Q}_{p})$ to explain the Wigner's Friend Paradox. Within this
framework, the geometry of the space $\mathbb{R\times Q}_{p}$ naturally causes
a collapse (localization) of the wavefunctions in space. This process does not
require observers or the interchange of information. In several
quantum-mechanical results, for instance, in the no-go theorems, the observers
are agents who can process information. They do not strictly need to be human,
but they must be capable of observing a system, recording the result in a
memory, and performing logical deductions. In the framework of QM on
$L^{2}(\mathbb{R}\times\mathbb{Q}_{p})$, what matters is not whether a system
is an agent, but whether it acts as an apparatus in the sense of Subsection
\ref{Section_Meas_I}. The paper was written in such a way that all the
calculations needed were essentially done in Sections \ref{Section_Meas_I} and
\ref{Section_Meas_II}. This fact also shows that the solution of the Wigner's
Friend paradox follows from the collapse mechanism.

We identify Wigner's friend with an apparatus $\mathcal{F}$, and use all the
notation and results given in Section \ref{Section_Meas_I}, with
$\mathcal{A}=\mathcal{F}$. In particular, the wavefunction describing the
quantum particle + Wigner's friend is
\[
\Psi_{\mathcal{S}+\mathcal{F}}\left(  t,x_{\infty},x_{p}\right)  =%
{\displaystyle\sum\limits_{m=1}^{\infty}}
{\displaystyle\sum\limits_{n=1}^{\infty}}
c_{n,m}\left(  t\right)  \psi_{n}\left(  x_{p}\right)  \theta_{m}\left(
x_{\infty}\right)  .
\]
We identify Wigner with an apparatus $\mathcal{W}$, and use the notation and
results given in Section \ref{Section_Meas_II}. More precisely, we identify
$\mathcal{S}_{2}=$ $\mathcal{W}$, \ and the system (quantum particle +
Wigner's friend + lab) corresponds to the quantum system $\mathcal{S}_{1}$.
Then the wavefunction of the system $\mathcal{W}+$ (quantum particle +
Wigner's friend + lab) is%

\begin{equation}
\Psi_{\mathcal{S}_{1}+\mathcal{W}}\left(  t,x_{\infty},x_{p},y_{\infty
}\right)  =%
{\displaystyle\sum\limits_{r=1}^{\infty}}
{\displaystyle\sum\limits_{m=1}^{\infty}}
{\displaystyle\sum\limits_{n=1}^{\infty}}
d_{n,m,r}\left(  t\right)  \psi_{n}\left(  x_{p}\right)  \omega_{r}\left(
y_{\infty}\right)  \theta_{m}\left(  x_{\infty}\right)  .
\label{wavefunction_S1_W}%
\end{equation}
It is relevant to mention here that we are using Remark \ref{Nota1}. More
precisely, we \ use the degree of freedom $y_{\infty}$ for $\mathcal{W}$, and
any extra degree of macroscopic freedom do not modify the analysis. \ For this
reason, we do not include a degree of freedom for the lab.

We denote by $T_{\text{Friend}}$, $T_{\text{Wigner }}$ the times at which the
measurements (interactions) end. In the classical paradox, $T_{\text{Friend}%
}<T_{\text{Wigner}}$. We discuss the paradox under the condition that the two
measurements do not occur at the same time.

\subsubsection{Scenario I: $T_{\text{Friend}}<T_{\text{Wigner}}$}

By the results of Section \ref{Section_Meas_I}, $\Psi_{\mathcal{S}%
+\mathcal{F}}\left(  T_{\text{Friend}},x_{\infty},x_{p}\right)  $ collapses in
space, and thus, the apparatus $\mathcal{F}$ produces a definite reading. Now,
starting out at the state%

\[
\widetilde{\Psi}_{\mathcal{S}+\mathcal{F}}^{B_{F}}\left(  x_{\infty}%
,x_{p}\right)  \Psi_{0}\left(  T_{\text{Friend}},y_{\infty}\right)  \in
L^{2}(\mathbb{R\times Q}_{p})%
{\textstyle\bigotimes}
L^{2}(\mathbb{R}),
\]
where $\widetilde{\Psi}_{\mathcal{S}+\mathcal{F}}^{B_{F}}$ is the state
(\ref{Lift}) of $\mathcal{S}+\mathcal{F}$ after the Friend's measurement, and
$B_{F}$ is the ball selected at $T_{\text{Friend}}$, the system $\mathcal{W}+$
(quantum particle + Wigner's friend + lab) is described by the wavefunction
(\ref{wavefunction_S1_W}), and by the results of Section \ref{Section_Meas_II}%
, $\Psi_{\mathcal{S}_{1}+\mathcal{W}}\left(  T_{\text{Wigner}},x_{\infty
},x_{p},y_{\infty}\right)  $ collapses in space, and thus, the apparatus
$\mathcal{W}$ produces a definite reading. There is no information exchange
between the apparatuses $\mathcal{W}$ and $\mathcal{F}$. As we mentioned in
Section \ref{Section_Meas_II}, the fact $\mathcal{W}$ (Wigner) gets a definite
reading requires the space discreteness hypothesis.

By the update rule (\ref{Update_rule}), in the form (\ref{Lift}), the
localized state of $\mathcal{S}+\mathcal{F}$ at $T_{\text{Friend}}$, an
element of $L^{2}(\mathbb{R})\otimes L^{2}(\mathbb{Q}_{p})$ supported in
$\mathcal{M}(B_{F})\times B_{F}$, is the initial state for the subsequent
evolution. The Friend's reading is a record that persists in the pointer
degree of freedom $x_{\infty}$ of $\mathcal{F}$, and Wigner's apparatus
interacts with this localized state. Since $\mathcal{F}$ is an apparatus in
the sense of Subsection \ref{Section_Meas_I}, $\mathcal{W}$ cannot undo the
Friend's measurement: there is no interference between the components removed
by the Friend's localization.

\subsubsection{Scenario II: $T_{\text{Friend}}>T_{\text{Wigner}}$}

By Section \ref{Section_Meas_II}, $\Psi_{\mathcal{S}_{1}+\mathcal{W}}\left(
T_{\text{Wigner}},x_{\infty},x_{p},y_{\infty}\right)  $ collapses in space,
and $\mathcal{W}$ produces a definite reading. Let $\Phi_{\mathcal{S}%
+\mathcal{F}}\left(  T_{\text{Wigner}},x_{\infty},x_{p}\right)  \in
L^{2}(\mathbb{R}\times\mathbb{Q}_{p})$ be the state of $\mathcal{S}%
+\mathcal{F}$ conditional on Wigner's reading, obtained from the update rule
(\ref{Update_rule_W})-(\ref{Lift_W}); in
general, it is not a product of a state of $\mathcal{S}$ and a state of
$\mathcal{F}$. Since $\mathcal{S}+\mathcal{F}$ may remain entangled with the
pointer of $\mathcal{W}$, this conditional state may be a mixed state (a
density operator); in this case, the results of Section \ref{Section_Meas_I}
apply to each of its pure components. Starting out at this state, the system
$\mathcal{S}+\mathcal{F}$ evolves in $L^{2}(\mathbb{R})\otimes L^{2}%
(\mathbb{Q}_{p})$. Since the results of Section \ref{Section_Meas_I} apply to
any initial state in this space,
\[
\Psi_{\mathcal{S}+\mathcal{F}}\left(  T_{\text{Friend}},\ \mathcal{M}%
(x_{p}),x_{p}\right)
\]
collapses (localizes) in space, so the apparatus $\mathcal{F}$ gives a
definite reading. Again, there is no information exchange between the
apparatuses $\mathcal{W}$ and $\mathcal{F}$.

\subsubsection{\label{Section_Persist}Order of the measurements and
persistence of records}

Let $B^{(1)},\ldots,B^{(s)}$ be the disjoint balls of radius $p^{-l_{F}}$
contained in the region scanned by the Friend; they correspond to the possible
readings of the Friend, and $p^{-l_{F}}$ is the resolution of the Friend's
measurement (Subsection \ref{Section_App}). By a \textit{component} of a state $\Psi$ we
mean a function of the form
\[
\Psi_{j}:=1_{B^{(j)}}(x_{p})\Psi,\qquad j=1,\ldots,s,
\]
i.e., the part of $\Psi$ supported, in the variable $x_{p}$, in the ball
$B^{(j)}$. The components have disjoint supports, and $\Psi=\sum_{j}\Psi_{j}$
when $\Psi$ is supported, in $x_{p}$, in the union of the balls $B^{(j)}$.

By a record of the Friend's reading, i.e., of the ball $B^{(j)}$ selected at
$T_{\text{Friend}}$, we mean the correlation, in the state, between the
component of the state supported in the ball $B^{(j)}$ and the corresponding
pointer positions $\mathcal{M}(B^{(j)})$. We say that the record
\textit{persists} on a time interval $[T_{\text{Friend}},t]$ if, during this
interval, the evolution does not recombine the component supported in
$B^{(j)}$ with the components supported in the balls corresponding to the
other possible readings of the Friend; otherwise, we say that the record is
\textit{undone}. When a record is undone, a later measurement can display
interference terms between the Friend's possible readings. After the update
(\ref{Update_rule}), the state of $\mathcal{S}+\mathcal{F}$ consists of the
component supported in $B^{(j)}$ alone; no other component remains with which
it could be recombined, so the record persists for every subsequent evolution:
it cannot be undone. After the measurement, the state
of $\mathcal{S}+\mathcal{F}$, including the Friend's pointer, evolves
according to the Schr\"{o}dinger equation, i.e., it follows the ordinary
evolution of a single state in standard QM. This evolution may change the
state, for instance, the free $p$-adic dynamics may move it out of the ball
$B^{(j)}$, but it does not erase the record, since no alternative component
remains with which it could interfere. These are properties of the state and
of the dynamics; they involve neither observers nor exchange of information.

We recall that, in this paper, the state is updated after each measurement:
by (\ref{Update_rule})-(\ref{Lift}) after the Friend's measurement, and by
(\ref{Update_rule_W})-(\ref{Lift_W}) after Wigner's measurement. Consequently,
whatever the order of the two measurements, each apparatus produces a definite
reading, and these readings persist as records in the pointer degrees of
freedom ($x_{\infty}$ for $\mathcal{F}$, and $y_{\infty}$ for $\mathcal{W}$),
and they cannot be undone. Thus, the framework does not eliminate memory: the
record is stored in the pointer and, at the microscopic level, in the $p$-adic
digits fixed by the localization. Indeed, the localization on the ball
$B=p^{l}a+p^{l}\mathbb{Z}_{p}$ fixes all the digits $x_{j}$, $j<l$, of the
$p$-adic coordinate. The update rule is essential here. Without the update,
the readings are still definite, but their records can be undone, in the sense
defined above, when the dynamics moves amplitude between the balls
corresponding to the different readings of the Friend; see Subsection
\ref{Section_Further} and Appendix F.

Whether the joint statistics of the readings of $\mathcal{F}$ and
$\mathcal{W}$ are the same in Scenarios I and II is a question that we do not
settle here. Each update is a restriction to a subset of the configuration
space followed by a normalization, and multiplications by characteristic
functions commute. However, in each scenario, a unitary evolution takes place
between the two updates, and the two updates act on different subsets of the
configuration space; as in the case of sequential measurements of
non-commuting observables in standard QM, the joint statistics may depend on
the order. This is an open problem.

\section{\label{Section_5}Consistency with No-Go Theorems}

The Wigner's Friend paradox has generated several no-go theorems. Each of them
asserts that a set of premises is incompatible with the predictions of QM, or
implies inequalities that QM violates. Such a theorem constrains our framework
only if each of its premises has a counterpart in the framework and holds
there; if a premise has no counterpart, or fails, the theorem does not apply.
We use this criterion throughout this section, and group the theorems
according to the notion on which they rely: agents, records of observed
events, and the Local Friendliness inequalities.

Two premises recur. The first, called the universal validity of QM, means, in
these theorems, that the Friend's measurement can be described by an outside
observer as a unitary interaction, so that it can, in principle, be reversed.
The second, freedom of choice, means that the super-observers (Wigner, or
Alice and Bob) choose their measurement settings independently of all the
variables relevant to the experiment that are prior to these choices. Our
framework does not model how the settings are chosen, so it imposes no
restriction on these choices: freedom of choice is compatible with the
framework, and the framework does not need to give it up.

\subsection{\label{Section_App}Apparatuses and quantum systems}

As stated in Subsection \ref{Section_Meas_I}, a system acts as an apparatus in
a measurement when it interacts with the quantum system $\mathcal{S}$ through
a map $\mathcal{M}$ from the microscopic space into the macroscopic one; then
the update rule (\ref{Update_rule}) applies. The radius $p^{-l}$ of the
scanned ball $B=p^{l}a+p^{l}\mathbb{Z}_{p}$ is the resolution of the
measurement, which fixes the digits $x_{j}$, $j<l$, of the $p$-adic
coordinate; the reading is definite at this resolution. A system that does not
interact with $\mathcal{S}$ through such a map behaves as a quantum system:
its evolution is unitary, and it can, in principle, be reversed, so that a
later measurement can display interference between the components of its state.

The no-go theorems discussed below are formulated for extended Wigner's Friend
scenarios, and some of them have been tested in experiments in which the
Friends are photonic degrees of freedom. Two kinds of Friends must therefore
be distinguished: the Friends of these experiments, which are quantum systems
in the above sense, and the Friend of Section \ref{Section_4}, who, like
Wigner, is an apparatus.

\begin{itemize}
\item \textit{The Friends of the photonic experiments.} In our framework, a
definite reading is produced by the localization on the graph of the map
$\mathcal{M}:\mathbb{Q}_{p}\rightarrow\mathbb{R}$ (Subsection
\ref{Section_Meas_I}). This requires an apparatus with a macroscopic
coordinate $x_{\infty}\in\mathbb{R}$, a quantum object with a microscopic
coordinate $x_{p}\in\mathbb{Q}_{p}$, and an interaction mediated by
$\mathcal{M}$; the update rule (\ref{Update_rule}) applies only in this
situation. In the experiments \cite{Proietti}, \cite{Bong et al}, the Friend
is a photonic degree of freedom that becomes entangled with $\mathcal{S}$
through a unitary gate; this coupling is not mediated by a map $\mathcal{M}$.
Consequently, no localization occurs, the update rule does not apply, and the
evolution is unitary: the interaction only entangles the Friend with
$\mathcal{S}$, and the joint state remains a superposition. Our framework
reproduces the predictions of standard QM (Subsection \ref{Section_LF}), and
the Friend's measurement can be reversed. In our framework, the only observed
events are the readings of pointers, i.e., localizations on the graph of a map
$\mathcal{M}$ (Subsection \ref{Section_Meas_I}). The no-go theorems use a
different notion: an observed event is the outcome of the Friend's
measurement, whatever the Friend is. The two notions agree for the Friend of
Section \ref{Section_4}, but not for a photonic Friend, which has no pointer:
its outcome is not an observed event in our framework, and the absoluteness of
observed events has no counterpart for it. This is how the framework
accommodates the observed violations of the Local Friendliness inequalities
with photonic Friends (Subsection \ref{Section_LFineq}).

\item \textit{The Friend of Section \ref{Section_4}.} The Friend is an
apparatus, as is Wigner. The Friend's reading is an observed event, and it
persists. But the update rule (\ref{Update_rule}) is neither unitary nor
linear, so Wigner cannot reverse the Friend's measurement: the universal
validity of QM fails in our framework.
\end{itemize}

\subsection{\label{Section_Agents}Agents: the Frauchiger--Renner theorem}

Frauchiger and Renner \cite{Frauchiger-Renner} consider agents: systems that
perform measurements, record the results, and reason about the conclusions of
other agents. They show that the following three assumptions are mutually
inconsistent: (Q) the universal validity of QM, including the unitary
description of the agents' laboratories; (C) an agent may use the conclusions
of other agents as his own; and (S) measurements have single definite
outcomes. In our framework, Wigner and his Friend are apparatuses, not agents:
they produce readings, but they do not reason about the readings of other
apparatuses. Assumption (C) has no counterpart in our framework, so the
theorem does not apply to it. In addition, assumption (S) holds, since the
localization produces a single definite reading.

\subsection{\label{Section_Records}Records of observed events}

In the no-go theorems, an observed event is the outcome of a measurement, as
registered by the system that performs it, and its record is the physical
trace of this outcome. In our framework, the observed events are the readings
of pointers (Subsection \ref{Section_App}), and their records are discussed in
Subsection \ref{Section_Persist}.

\textit{Brukner} \cite{Brukner} shows that the universal validity of QM,
locality, freedom of choice, and the existence of observer-independent facts
are mutually incompatible. For apparatus Friends, the universal validity of QM
fails in our framework. For Friends that are quantum systems, the interaction
with $\mathcal{S}$ produces no reading of a pointer, only an entangled
superposition; hence, the Friend's outcome is not an observed event in our
framework, and the existence of observer-independent facts has no counterpart
for it. In both cases, the theorem does not apply, and freedom of choice need
not be given up.

\textit{Gu\'{e}rin et al.} \cite{Guerin et al} show that the linearity of QM,
applied to the Friend's measurement as described by Wigner, and the persistent
reality of the Friend's records at two different times are mutually
incompatible. In our framework, the records persist (Subsection
\ref{Section_Persist}), and linearity fails at the update (\ref{Update_rule});
the theorem does not apply.

\textit{Ormrod and Barrett} \cite{Ormrod-Barrett} show, without inequalities
or modal logic, that the absoluteness of observed events is incompatible with
the unitary description of measurements together with the assumption that
quantum theory applies in any inertial reference frame. In our framework, the
readings are absolute, but the unitary description of measurements fails at
the update, and the framework is not Lorentz invariant at the microscopic
level; the theorem does not apply.

\subsection{\label{Section_LFineq}The Local Friendliness inequalities}

Bong et al. \cite{Bong et al} consider two Friends, each of whom measures one
of a pair of entangled systems inside a closed laboratory, and two
super-observers, Alice and Bob, each of whom either asks the Friend for the
outcome or reverses the Friend's measurement and measures in a different
basis. \textit{Local Friendliness} is the conjunction of three assumptions:
\textit{absoluteness of observed events} (every observed event, including the
Friend's outcome, is a single event, not relative to anything or anyone);
\textit{no-superdeterminism}, which is the freedom of choice described above;
and \textit{locality} (a choice of setting cannot influence outcomes in
regions that are space-like separated from it). These assumptions imply a set
of inequalities for the correlations observed by Alice and Bob, the
\textit{Local Friendliness inequalities}. Bong et al. show that QM violates
these inequalities if quantum evolution is controllable on the scale of an
observer, i.e., if the universal validity of QM holds for the Friends.
Violations have been observed with photonic Friends \cite{Proietti},
\cite{Bong et al}.

The Local Friendliness inequalities concern measured correlations, so any
framework describing these experiments must specify which of the three
assumptions fails in them. In our framework, the readings of the apparatuses
are absolute, no-superdeterminism is compatible, as explained above, and
locality fails, since QM on $L^{2}(\mathbb{R}\times\mathbb{Q}_{p})$ is
non-local \cite{Zuniga-2026}-\cite{Zuniga-QM-2}, \cite{Zuniga-PhA}. Hence the
theorem of Bong et al. does not constrain our framework: a violation of the
Local Friendliness inequalities is not excluded. This consistency does not, by
itself, decide whether the Friend's measurement can be described by an outside
observer as a unitary interaction; the answer depends on the kind of Friend,
and it follows from the criterion of Subsection \ref{Section_App}.

\begin{itemize}
\item \textit{Friends that are quantum systems.} The Friend's measurement is a
unitary interaction, which Alice and Bob can reverse, so the universal
validity of QM holds for these Friends, and the framework reproduces the
observed violation (Subsection \ref{Section_LF}). No observed event occurs at
the Friend, so the absoluteness of observed events has no counterpart for the
Friend's outcome.

\item \textit{Friends that are apparatuses.} The Friend's measurement is
described by the update rule (\ref{Update_rule}), which is neither unitary
nor linear, so Alice and Bob cannot reverse it: the condition under which Bong
et al. derive the violation, namely that quantum evolution is controllable on
the scale of the Friend, fails in our framework. The readings are absolute,
and the framework does not predict a violation (Subsection \ref{Section_Exp}).
\end{itemize}

\subsection{\label{Section_LF}Local Friendliness scenario with Friends that
are quantum systems}

We now consider the scenario of Bong et al. described in Subsection
\ref{Section_LFineq}. When the Friends are quantum systems, as in the
experiments \cite{Proietti}, \cite{Bong et al}, all the operations involved
are unitary operators and projective measurements on a finite-dimensional
Hilbert space $\mathbb{C}^{N_{1}}\otimes\cdots\otimes\mathbb{C}^{N_{k}}$, one
factor for each system. By Appendix B, each factor $\mathbb{C}^{N_{j}}$ can be
identified with a subspace $\chi_{N_{j}}(\mathbb{Z}_{p})\subset L^{2}%
(\mathbb{Z}_{p})$, and the scenario can be implemented in $L^{2}%
(\mathbb{Z}_{p}^{k})$ by an isometry $V$ that preserves the tensor product
structure: each unitary operator $U$ is mapped to $VUV^{\ast}+\left(
1-VV^{\ast}\right)  $, and each orthogonal projection $P$ to $VPV^{\ast}$; see
Appendix E. Consequently:

\begin{enumerate}
\item All joint probabilities coincide with those of standard QM; in
particular, the Local Friendliness inequalities are violated by the amount
predicted by QM \cite{Bong et al}, in agreement with the experiments.

\item The operations of Alice act on her tensor factor only; hence, as in
standard QM, the marginal distributions of Alice do not depend on the settings
of Bob, and vice versa: the statistics are no-signalling.
\end{enumerate}

When the Friends are apparatuses, Alice and Bob cannot reverse the Friends'
measurements. The analysis of the correlations in this regime, including the
question of no-signalling at the level of observable statistics in view of the
superluminal transition probabilities found in \cite{Zuniga-PhA}, requires
substantial new work and is left as an open problem.

\subsection{\label{Section_Exp}Experimental consequences}

The previous discussion leads to a prediction. In extended Wigner's Friend
experiments, the framework predicts that violations of the Local Friendliness
inequalities occur when the Friends are quantum systems, as observed in
\cite{Proietti}, \cite{Bong et al}, and cease when the Friends are apparatuses
in the sense of Subsection \ref{Section_App}. Improved isolation of the
Friends' laboratories does not restore the violations, because the
localization is an update of the state and not an effect of decoherence.
Unitary QM predicts that the violations persist, in principle, for Friends of
any size, provided that the laboratories are sufficiently isolated. An
extended Wigner's Friend experiment with Friends of increasing size and
complexity (see \cite{Wiseman} for a related proposal) would therefore test
the framework; the size at which the violations cease would be a measurement
of the boundary between the microscopic and macroscopic spaces that the
framework posits. The present model does not yet give a quantitative value for
this boundary, since it depends on the map $\mathcal{M}$ and on the depth of
the measurement; determining it is a problem for future work.

\subsection{Relativistic no-go results}

Extensions of the Wigner's Friend scenarios to relativistic and
field-theoretic settings \cite{Durham}-\cite{Wiseman}, in particular those of
Allam and Matzkin \cite{Allam -Matzkin}-\cite{Allam-Matzkin-2}, who show that
the dependence of the state update on the reference frame leads to
inconsistent accounts of outcomes, rely on observers in different inertial
frames who describe the same measurements. Relativity requires such observers;
our framework requires neither observers nor intelligent agents. The collapse
mechanism operates independently of any observer and does not appeal to
Lorentz invariance, since the update (\ref{Update_rule}) involves the $p$-adic
coordinate, whose sector $\left\{  \beta\right\}  \times\mathbb{Q}_{p}$ has no
Lorentz symmetry. The premises of these results have no counterpart in our
framework, so they do not apply to it.

\section{\label{Section_6}Conclusions}

In this paper we have proposed a resolution of the original Wigner's Friend
paradox within the framework of quantum mechanics on the hybrid space
$\mathbb{R}\times\mathbb{Q}_{p}$, introduced in \cite{Zuniga-2026}. The
resolution rests on the following interconnected ideas.

\textbf{Collapse as localization, with a physical specification of
measurement}. In QM on $L^{2}(\mathbb{R}\times\mathbb{Q}_{p})$, systems evolve
according to the Schr\"{o}dinger equation with non-local Hamiltonians between
measurements. A measurement is an interaction between a quantum system and an
apparatus, mediated by a continuous map $\mathcal{M}$ from the microscopic
space into the macroscopic one. During this interaction, the state is
localized on the graph of $\mathcal{M}$ over a ball $B$ of $\mathbb{Q}_{p}$
selected by the Born rule, according to the update rule (\ref{Update_rule}),
which is the standard projection rule of the Dirac--von Neumann formalism. The
map $\mathcal{M}$ appears as a parameter: it does not change the Dirac--von
Neumann formulation, but it changes the interpretation of quantum
measurements, which are interactions that correlate the pointer with the
quantum object through $\mathcal{M}$. What the framework adds is thus a
physical specification of when the projection rule applies and of the
projection involved; the ultrametric property of $\mathbb{Q}_{p}$ (two balls
are either disjoint or nested) gives the explicit form of the localized state.
Neither observers, nor new physical constants, nor a modification of the
Schr\"{o}dinger equation are required. The localization, and hence the
definiteness of each reading, does not depend on the update rule; the update
rule enters only as the initial condition for the Schr\"{o}dinger evolution
that follows a measurement.

\textbf{No privileged observers}. Collapse requires no observers, no conscious
agents, and no communication between subsystems. Both Wigner and his Friend
are modeled as apparatuses interacting with quantum systems, each governed by
the same localization mechanism. In the standard paradox, the conflict arises
because Friend and Wigner appear to assign incompatible states to the same
system at the same time. In the present framework there is no conflict: each
apparatus produces a definite reading, which persists as a record, without
requiring the other to exist.

\textbf{Apparatuses and quantum systems}. The framework provides a physical
criterion that distinguishes apparatuses from quantum systems: a system acts
as an apparatus when it interacts with the quantum system through a map
$\mathcal{M}$; the radius of the scanned ball is the resolution of the measurement.

\textbf{Consistency with no-go theorems}. The framework is consistent with the
principal no-go theorems associated with extended Wigner's Friend scenarios
(Frauchiger--Renner, Brukner, Bong et al., Gu\'{e}rin et al.,
Ormrod--Barrett). For the theorem of Frauchiger and Renner, which involves
agents reasoning about the conclusions of other agents, the consistency is
immediate, since the framework contains apparatuses but no agents. For the
other theorems, it follows from the criterion that distinguishes apparatuses
from quantum systems. For Friends that are quantum systems, the evolution is
unitary and the framework reproduces the quantum correlations, including the
observed violations of the Local Friendliness inequalities, with no-signalling
statistics; but no observed event occurs at the Friend. For apparatus Friends,
the Friend's reading is an absolute and persistent observed event, but the
Friend's measurement is not unitary, and Wigner cannot reverse it. The
framework keeps the absoluteness of the readings, imposes no restriction on
the choice of the measurement settings, and it is non-local.

\textbf{A testable prediction}. The framework predicts that violations of the
Local Friendliness inequalities cease when the Friends are apparatuses, and
that improved isolation does not restore them, whereas unitary QM predicts
that they persist in principle. Extended Wigner's Friend experiments with
Friends of increasing size would test this prediction.

\textbf{Realism}. By realism we mean the existence of a physical reality
independent of any observer. In our framework, the observer-independent
entities are the readings of the apparatuses and the probability measures
defined by the wavefunctions. We do not claim value definiteness, i.e., that
all observables possess definite values before measurement.

\textbf{The role of the extra coordinate}. The analysis clarifies why
resolving the measurement problem for Wigner himself --- as opposed to his
Friend --- requires the larger space $\mathbb{R}\times\mathbb{Q}_{p}$ rather
than $\mathbb{Q}_{p}$ alone. In the Friend's measurement, the measured quantum
system has microscopic space $\mathbb{Q}_{p}$, while the Friend, as an
apparatus, has macroscopic space $\mathbb{R}$. Wigner's measurement treats the
system formed by the quantum particle and the Friend as the measured system;
its configuration space is $\mathbb{R}\times\mathbb{Q}_{p}$. The transition
from one model to the other corresponds to adding an extra, non-Archimedean,
coordinate. This step is not merely a formal convenience: it reflects the
different physical characters of the two measurements.

\textbf{Some open questions remain}. (1) The choice of the space $\mathcal{X}$
and of the map $\mathcal{M}:\mathcal{X}\rightarrow\mathbb{R}$ is a structural
assumption of the model. A natural `philosophical' question is the existence
of a universal model $\mathcal{X}$ for the space at the microscopic level.
Here, following \cite{Zuniga-2026} and \cite{Berezin et al}, we argue that in
QM, the selection of a Hilbert space also includes the selection of a
configuration space $\mathcal{X}$, and that the selection is made considering
the particular phenomena to be modeled. (2) The alternative reading, in which
the localization is only a projection while the global state evolves
unitarily, requires showing that Wigner's measurement removes the interference
terms. (3) The dependence of the joint statistics of Wigner and his Friend on
the order of the measurements. (4) The correlations in extended Wigner's
Friend scenarios with apparatus Friends, and the no-signalling property at the
level of observable statistics, in view of the superluminal transition
probabilities of \cite{Zuniga-PhA}. (5) A relativistic formulation of the
update rule, and a quantitative value for the boundary between microscopic and
macroscopic systems.

\bigskip

\textbf{Acknowledgements}

The author was partially supported by the Debnath Endowment.

\textbf{Use of Generative AI and AI-Assisted Technologies in the Writing
Process}

During the revising process of this work, the author used Claude (Anthropic;
model Claude Opus 5.5) to check some calculations, and to improve the clarity
and readability of the language of select sections. After using this tool, the
author reviewed and edited the content as needed and takes full responsibility
for the publication.

\textbf{Conflict of Interest}

The author declares no conflict of interest.

\textbf{Data Availability Statement}

No data were utilized in this theoretical paper.

\textbf{Keywords}

$p$-adic quantum mechanics, $p$-adic numbers, The Wigner's friend paradox,
Planck length.

\section{\label{Appendix A}Appendix A: Basic aspects of the $p$-adic Analysis}

In this appendix, we collect some basic results from the $p$-adic analysis.
For a detailed exposition on $p$-adic analysis, the reader may consult
\cite{V-V-Z}-\cite{Zuniga-Textbook}. Our presentation here is based on the
book \cite{Zuniga-Textbook}.

\subsection{The field of $p$-adic numbers}

Let $p$ be a fixed prime number. The field of $p-$adic numbers $\mathbb{Q}%
_{p}$ is defined as the completion of the field of rational numbers
$\mathbb{Q}$ with respect to the $p-$adic norm $|\cdot|_{p}$, which is defined
as
\[
|x|_{p}=%
\begin{cases}
0 & \text{if }x=0\\
p^{-\gamma} & \text{if }x=p^{\gamma}\dfrac{a}{b},
\end{cases}
\]
where $a$ and $b$ are integers coprime with $p$. The integer $\gamma
=ord_{p}(x):=ord(x)$, with $ord(0):=+\infty$, is called the\textit{\ }$p-$adic
order of $x$. The metric space $\left(  \mathbb{Q}_{p},|\cdot|_{p}\right)  $
is a complete ultrametric space. As a topological space $\mathbb{Q}_{p}$\ is
homeomorphic to a Cantor-like subset of the real line, see, e.g.,
\cite{V-V-Z}, \cite{Alberio et al}.

Any $p-$adic number $x\neq0$ has a unique expansion of the form
\[
x=p^{ord(x)}\sum_{j=0}^{\infty}x_{j}p^{j},
\]
where $x_{j}\in\{0,1,2,\dots,p-1\}$ and $x_{0}\neq0$. By using this expansion,
we define \textit{the fractional part }$\{x\}_{p}$\textit{ of }$x\in
\mathbb{Q}_{p}$ as the rational number
\[
\{x\}_{p}=%
\begin{cases}
0 & \text{if }x=0\text{ or }ord(x)\geq0\\
p^{ord(x)}\sum_{j=0}^{-ord(x)-1}x_{j}p^{j} & \text{if }ord(x)<0.
\end{cases}
\]
In addition, any $x\in\mathbb{Q}_{p}\smallsetminus\left\{  0\right\}  $
can be represented uniquely as $x=p^{ord(x)}v$, where $\left\vert v\right\vert
_{p}=1$.

\subsection{Topology of $\mathbb{Q}_{p}$}

For $r\in\mathbb{Z}$, denote by $B_{r}(a)=\{x\in\mathbb{Q}_{p};|x-a|_{p}\leq
p^{r}\}$ the ball of radius $p^{r}$ with center at $a\in\mathbb{Q}_{p}$, and
take $B_{r}(0):=B_{r}$. The ball $B_{0}=\mathbb{Z}_{p}$ is the ring of\textit{
}$p-$adic integers. Notice that $B_{r}(a)=a+p^{-r}\mathbb{Z}_{p}$, where
\[
\mathbb{Z}_{p}=\left\{  x\in\mathbb{Q}_{p};\left\vert x\right\vert _{p}%
\leq1\right\}  .
\]
We denote by $\Omega\left(  p^{-r}\left\vert x-a\right\vert _{p}\right)  $ the
characteristic function of the ball $B_{r}(a)$; for instance, $\Omega\left(
\left\vert x\right\vert _{p}\right)  $ is the characteristic function of
$\mathbb{Z}_{p}$. For a general set $A$, we denote by $1_{A}$ its
characteristic function.
We also denote by $S_{r}(a)=\{x\in\mathbb{Q}_{p};|x-a|_{p}=p^{r}\}$ the sphere
of radius\textit{ }$p^{r}$ with center at $a\in\mathbb{Q}_{p}$, and take
$S_{r}(0):=S_{r}$. We notice that $S_{0}=\mathbb{Z}_{p}^{\times}$ (the group
of units of $\mathbb{Z}_{p}$). The balls and spheres are both open and closed
subsets in $\mathbb{Q}_{p}$. In addition, two balls in $\mathbb{Q}_{p}$
are either disjoint or one is contained in the other.

As a topological space $\left(  \mathbb{Q}_{p},|\cdot|_{p}\right)  $ is
totally disconnected, i.e., the only connected \ subsets of $\mathbb{Q}_{p}$
are the empty set and the points. A subset of $\mathbb{Q}_{p}$ is compact if
and only if it is closed and bounded in $\mathbb{Q}_{p}$, see, e.g.,
\cite[Section 1.3]{V-V-Z}, or \cite[Section 1.8]{Alberio et al}. The balls and
spheres are compact subsets. Thus $\left(  \mathbb{Q}_{p},|\cdot|_{p}\right)
$ is a locally compact topological space.

\subsection{Additive characters}

Set $\chi_{p}(y):=\exp(2\pi i\{y\}_{p})$ for $y\in\mathbb{Q}_{p}$. The map
$\chi_{p}(\cdot)$ is an additive character on $\mathbb{Q}_{p}$, i.e., a
continuous map from $\left(  \mathbb{Q}_{p},+\right)  $ into $S$, the unit
circle considered as a multiplicative group, satisfying $\chi_{p}(x_{0}%
+x_{1})=\chi_{p}(x_{0})\chi_{p}(x_{1})$, $x_{0},x_{1}\in\mathbb{Q}_{p}$; see,
e.g., \cite[Section 2.3]{Alberio et al}.

\subsection{The Haar measure}

Since $(\mathbb{Q}_{p},+)$ is a locally compact topological group, there
exists a Haar measure $dx$, which is invariant under translations, i.e.,
$d(x+a)=dx$. If we normalize this measure by the condition $\int
_{\mathbb{Z}_{p}}dx=1$, then $dx$ is unique.

\subsection{The Bruhat-Schwartz space}

A complex-valued function $\varphi$ defined on $\mathbb{Q}_{p}$ is
\textit{called locally constant} if for any $x\in\mathbb{Q}_{p}$ there exist
an integer $l(x)\in\mathbb{Z}$ such that%
\begin{equation}
\varphi(x+x^{\prime})=\varphi(x)\text{ for any }x^{\prime}\in B_{l(x)}.
\label{local_constancy}%
\end{equation}
A function $\varphi:\mathbb{Q}_{p}\rightarrow\mathbb{C}$ is called a
Bruhat-Schwartz function\textit{ }(or a test function) if it is locally
constant with compact support. Any test function can be represented as a
linear combination, with complex coefficients, of characteristic functions of
balls. The $\mathbb{C}$-vector space of Bruhat-Schwartz functions is denoted
by $\mathcal{D}(\mathbb{Q}_{p})$. For $\varphi\in\mathcal{D}(\mathbb{Q}_{p})$,
the largest number $l=l(\varphi)$ satisfying (\ref{local_constancy}) is called
the exponent of local constancy (or the parameter of constancy) of $\varphi$.

\subsection{$L^{\rho}$ spaces}

Given an open subset $U\subset\mathbb{Q}_{p}$ (for instance $U=\mathbb{Z}_{p}%
$, $\mathbb{Q}_{p}$), $\mathcal{D}(U)$ denotes the $\mathbb{C}$-vector space
of test functions with supports contained in $U$. For $\rho\in\lbrack
1,\infty)$, we denote by $L^{\rho}\left(  U\right)  =L^{\rho}\left(
U,x\right)  $, the $\mathbb{C}-$vector space of all the complex valued
functions $g$ satisfying
\[
\left\Vert g\right\Vert _{\rho}=\left(  \text{ }%
{\displaystyle\int\limits_{U}}
\left\vert g\left(  x\right)  \right\vert ^{\rho}dx\right)  ^{\frac{1}{\rho}%
}<\infty,
\]
where $dx$ is the normalized Haar measure on $\left(  \mathbb{Q}_{p},+\right)
$. Furthermore, $\mathcal{D}(U)$ is dense in $L^{\rho}\left(  U\right)  $ for
$1\leq\rho<\infty$, see, e.g., \cite[Section 4.3]{Alberio et al}.

\section{\label{Appendix B}Appendix B: QM on $\mathbb{C}^{N}$ is QM on
$L^{2}\left(  \mathbb{Z}_{p}\right)  $}

The results of this appendix are a reformulation of some results already
presented in \cite{Zuniga-2026}. Set $G_{l}=\left\{  I=I_{0}+\ldots
+I_{l-1}p^{l-1}\text{, with }I_{k}\in\left\{  0,\ldots,p-1\right\}  \right\}
$, and denote by $\Omega\left(  p^{l}\left\vert x-I\right\vert _{p}\right)  $
the characteristic function of the ball $I+p^{l}\mathbb{Z}_{p}$. The
functions $\left\{  p^{\frac{l}{2}}\Omega\left(  p^{l}\left\vert
x-I\right\vert _{p}\right)  \right\}  _{I\in G_{l}}$ form an orthonormal
basis, with respect to the inner product%
\[
\left\langle \varphi,\phi\right\rangle =%
{\displaystyle\int\limits_{\mathbb{Z}_{p}}}
\varphi\left(  x\right)  \overline{\phi}\left(  x\right)  dx.
\]
Indeed,%
\begin{gather*}
\left\langle p^{\frac{l}{2}}\Omega\left(  p^{l}\left\vert x-I\right\vert
_{p}\right)  ,p^{\frac{l}{2}}\Omega\left(  p^{l}\left\vert x-J\right\vert
_{p}\right)  \right\rangle =p^{l}%
{\displaystyle\int\limits_{\mathbb{Z}_{p}}}
\Omega\left(  p^{l}\left\vert x-I\right\vert _{p}\right)  \overline
{\Omega\left(  p^{l}\left\vert x-J\right\vert _{p}\right)  }dx\\
=p^{l}%
{\displaystyle\int\limits_{\left(  I+p^{l}\mathbb{Z}_{p}\right)  \cap\left(
J+p^{l}\mathbb{Z}_{p}\right)  }}
dx=p^{l}\delta_{I,J}%
{\displaystyle\int\limits_{I+p^{l}\mathbb{Z}_{p}}}
dx=\delta_{I,J},
\end{gather*}
where $\delta_{I,J}$ denotes the Kronecker delta. We denote by $\mathcal{D}%
_{l}(\mathbb{Z}_{p})$ the complex Hilbert space spanned by $\left\{
p^{\frac{l}{2}}\Omega\left(  p^{l}\left\vert x-I\right\vert _{p}\right)
\right\}  _{I\in G_{l}}$ with dimension $\#G_{l}=p^{l}$.

We select a subset $G_{l}^{0}\subseteq G_{l}$, with cardinality $\#G_{l}%
^{0}=N$, and denote by $\mathcal{\chi}_{N}(\mathbb{Z}_{p})$ the $N$%
-dimensional Hilbert space spanned by%
\begin{equation}
\left\{  p^{\frac{l}{2}}\Omega\left(  p^{l}\left\vert x-I\right\vert
_{p}\right)  ;I\in G_{l}^{0}\right\}  \subset\mathcal{D}_{l}(\mathbb{Z}%
_{p}).\label{basis}%
\end{equation}
Then, $\mathcal{\chi}_{N}(\mathbb{Z}_{p})$ and $\mathbb{C}^{N}$ are isomorphic
as Hilbert spaces ($\mathcal{\chi}_{N}(\mathbb{Z}_{p})\simeq$ $\mathbb{C}^{N}%
$). Notice that the isomorphism exists for any $p$ and $l$ such that $N\leq
p^{l}$.

We identify $p^{\frac{l}{2}}\Omega\left(  p^{l}\left\vert x-I\right\vert
_{p}\right)  $ with $e_{I}$, where $\left\{  e_{I}\right\}  _{I\in G_{l}^{0}}$
is the canonical basis of $\mathbb{C}^{N}$. Let
\[
\left[  H_{J,K}\right]  _{1\leq J,K\leq N}=\left[  H_{J,K}\right]  _{J,K\in
G_{l}^{0}}%
\]
be a Hermitian matrix, i.e., a Hamiltonian on $\mathcal{\chi}_{N}%
(\mathbb{Z}_{p})\simeq$ $\mathbb{C}^{N}$. We attach to this matrix the kernel%
\[
h(x,y):=p^{l}%
{\displaystyle\sum\limits_{J\in G_{l}^{0}}}
\text{ }%
{\displaystyle\sum\limits_{K\in G_{l}^{0}}}
H_{J,K}\Omega\left(  p^{l}\left\vert x-J\right\vert _{p}\right)  \Omega\left(
p^{l}\left\vert y-K\right\vert _{p}\right)  ,
\]
for $x,y\in\mathbb{Z}_{p}$, and the linear operator%
\begin{equation}
\varphi\left(  x\right)  \rightarrow\boldsymbol{H}\varphi\left(  x\right)  =%
{\displaystyle\int\limits_{\mathbb{Z}_{p}}}
h(x,y)\varphi\left(  y\right)  dy,\label{Integral}%
\end{equation}
where $\varphi\left(  y\right)  =\sum_{K\in G_{l}^{0}}c_{K}\Omega\left(
p^{l}\left\vert y-K\right\vert _{p}\right)  \in\mathcal{\chi}_{N}%
(\mathbb{Z}_{p})$. This operator can be extended to the space $\mathcal{C}%
(\mathbb{Z}_{p})$ of continuous functions defined on $\mathbb{Z}_{p}$, i.e.,
$\boldsymbol{H}:\mathcal{C}(\mathbb{Z}_{p})\rightarrow\mathcal{C}%
(\mathbb{Z}_{p})$ is a well-defined linear operator. Now, we use the fact that
$\mathcal{C}(\mathbb{Z}_{p})$ is dense in $L^{2}(\mathbb{Z}_{p})$, and the
estimation$\mathcal{\ }$
\[
\left\Vert \boldsymbol{H}\varphi\right\Vert _{L^{2}(\mathbb{Z}_{p})}%
\leq\left\Vert h\right\Vert _{L^{2}(\mathbb{Z}_{p}\times\mathbb{Z}_{p}%
)}\left\Vert \varphi\right\Vert _{L^{2}(\mathbb{Z}_{p})}%
\]
to conclude \ that $\boldsymbol{H}$ has unique extension as a linear, bounded
operator on $L^{2}(\mathbb{Z}_{p})$, that we denote again by $\boldsymbol{H}$.
Notice that this operator is non-local, and that by construction
$\mathcal{\chi}_{N}(\mathbb{Z}_{p})$ is invariant under $\boldsymbol{H}$, and
the restriction $\boldsymbol{H:}$ $\mathcal{\chi}_{N}(\mathbb{Z}%
_{p})\rightarrow\mathcal{\chi}_{N}(\mathbb{Z}_{p})$ is represented by the
matrix $\left[  H_{J,K}\right]  _{1\leq J,K\leq N}$. It is not difficult to
verify that $\boldsymbol{H}$ is symmetric, and consequently self-adjoint. In
conclusion,%
\[
i\frac{\partial}{\partial t}\Psi\left(  x,t\right)  =\boldsymbol{H}\Psi\left(
x,t\right)  \text{, }x\in\mathbb{Z}_{p}\text{, }t\geq0,
\]
is a continuous Schr\"{o}dinger equation attached to matrix $\left[
H_{J,K}\right]  _{1\leq J,K\leq N}$.

\section{Appendix C: the Vladimirov operator and orthonormal basis for
$L^{2}(\mathbb{Q}_{p})$}

The simplest Hamiltonian is $\boldsymbol{D}^{\alpha}$, $\alpha>0$, the
Vladimirov fractional derivative, which is defined as%
\begin{equation}
\boldsymbol{D}^{\alpha}\varphi\left(  x\right)  =\frac{1-p^{\alpha}%
}{1-p^{-\alpha-1}}%
{\displaystyle\int\limits_{\mathbb{Q}_{p}}}
\frac{\varphi(z)-\varphi(x)}{|z-x|_{p}^{\alpha+1}}%
\,dz,\label{Vladimirov-Taibleson-Derivative}%
\end{equation}
for $\varphi$ a locally constant function with compact support; see, e.g.,
\cite[Chapter 2]{Zuniga-Textbook}. To see the non-local nature of this
operator, we take $\varphi\left(  x\right)  =1$ if $|x|_{p}\leq1$, otherwise
$\varphi\left(  x\right)  =0$, then%
\begin{align*}
\boldsymbol{D}^{\alpha}\varphi\left(  x\right)   &  =\frac{1-p^{\alpha}%
}{1-p^{-\alpha-1}}\left\{
{\displaystyle\int\limits_{|z|_{p}\leq1}}
\frac{\varphi(z)-\varphi(x)}{|z-x|_{p}^{\alpha+1}}\,dz+%
{\displaystyle\int\limits_{|z|_{p}>1}}
\frac{\varphi(z)-\varphi(x)}{|z-x|_{p}^{\alpha+1}}\,dz\right\}  \\
&  =\left\{
\begin{array}
[c]{lll}%
-\frac{1-p^{\alpha}}{1-p^{-\alpha-1}}\left(  \text{ }%
{\displaystyle\int\limits_{|z|_{p}>1}}
\frac{dz}{|z|_{p}^{\alpha+1}}\,\right)   & \text{if} & |x|_{p}\leq1\\
&  & \\
\frac{1-p^{\alpha}}{1-p^{-\alpha-1}}\frac{1}{|x|_{p}^{\alpha+1}} & \text{if} &
|x|_{p}>1.
\end{array}
\right.
\end{align*}

\subsection{An orthonormal basis for $L^{2}(\mathbb{Q}_{p})$}

We set%
\[
\mathbb{Q}_{p}/\mathbb{Z}_{p}=\left\{  \sum_{j=-1}^{-m}x_{j}p^{j};\text{for
some }m>0\right\}  .
\]
For $b\in\mathbb{Q}_{p}/\mathbb{Z}_{p}$, $r\in\mathbb{Z}$, we denote by
$\Omega\left(  \left\vert p^{r}x_{p}-b\right\vert _{p}\right)  $ the
characteristic function of the ball $bp^{-r}+p^{-r}\mathbb{Z}_{p}$. We now
define
\[
\psi_{rbk}\left(  x_{p}\right)  =p^{\frac{-r}{2}}\chi_{p}(p^{-1}k\left(
p^{r}x_{p}-b\right)  )\Omega\left(  \left\vert p^{r}x_{p}-b\right\vert
_{p}\right)  ,
\]
where $r\in\mathbb{Z}$, $k\in\{1,\dots,p-1\}$, and $b\in\mathbb{Q}%
_{p}/\mathbb{Z}_{p}$.

Then, $\left\{  \psi_{rbk}\left(  x_{p}\right)  \right\}  _{rbk}$ forms a
complete orthonormal basis of $L^{2}(\mathbb{Q}_{p})$, and
\[
\boldsymbol{D}^{\alpha}\psi_{rbk}\left(  x\right)  =p^{\left(  1-r\right)
\alpha}\psi_{rbk}\left(  x\right)  \text{, for any }r,b,k;
\]
see, e.g., \cite[Theorems 9.4.5 and 8.9.3]{Alberio et al}, or \cite[Theorem
3.3]{KKZuniga}.

We now compute the restriction of $\psi_{rbk}\left(  x_{p}\right)  $ to the
ball $B=p^{l}a+p^{l}\mathbb{Z}_{p}$:%
\begin{gather}
\Omega\left(  p^{l}\left\vert x_{p}-p^{l}a\right\vert _{p}\right)  \psi
_{rbk}\left(  x_{p}\right)  =\label{TableA}\\
\left\{
\begin{array}
[c]{ll}%
\psi_{rbk}\left(  x_{p}\right)  & \text{if }bp^{-r}-p^{l}a\in p^{l}%
\mathbb{Z}_{p}\text{, }r\leq-l\\
& \\
p^{\frac{-r}{2}}\chi_{p}\left(  p^{-1}k\left(  p^{r+l}a-b\right)  \right)
\Omega\left(  p^{l}\left\vert x_{p}-p^{l}a\right\vert _{p}\right)  & \text{if
}bp^{-r}-p^{l}a\in p^{-r}\mathbb{Z}_{p}\text{, }r\geq-l+1\\
& \\
0 & \text{if }bp^{-r}-p^{l}a\notin p^{-r}\mathbb{Z}_{p}\text{, }r\geq-l+1,
\end{array}
\right. \nonumber
\end{gather}
where $\Omega\left(  p^{l}\left\vert x_{p}-p^{l}a\right\vert _{p}\right)  $
denotes the characteristic function of $p^{l}a+p^{l}\mathbb{Z}_{p}$. The above
calculation has appeared in several publications; see, e.g., \cite[Table
4.4]{Zuniga-PhA}, and the references therein.

\section{\label{Appendix D}Appendix D: A toy model: the particle in a box}

The collapse mechanism introduced in Section \ref{Section_2} is a `kind of
mathematical result' showing that the wavefunction that appears in the
measurement process has compact support, which implies that the pointer gives
a definite reading. In this appendix, we study a toy model of measuring the
energy levels of a particle in a box. The detailed estimates are given in the
last subsection of this appendix.

\subsection{The collapse in a continuous space}

In this subsection, we use some results from \cite[Chapter 3]{Norsen} and from
\cite{Zuniga-Mayes}. In this last work, the $p$-adic Schr\"{o}dinger equation
for a particle in a box is studied. We work on the Hilbert space
$L^{2}(\mathbb{X})\otimes L^{2}(\mathbb{R})\simeq L^{2}(\mathbb{X}%
\times\mathbb{R})$, where $\mathbb{X}$ denotes $\mathbb{R}$ or $\mathbb{Q}%
_{p}$. We suppose that $\psi_{n}$, $n=1,2,\ldots$ is an orthonormal basis of
$L^{2}(\mathbb{X})$. The \ particle in the box start out in the state%
\[
\Theta_{0}\left(  x\right)  =%
{\displaystyle\sum\limits_{n=1}^{\infty}}
c_{n}\psi_{n}\left(  x\right)  \text{, \ }x\in\mathbb{X}\text{.}%
\]
The pointer in its ready position is described by a Gaussian wave packet
centered on the position $y_{0}\in\mathbb{R}$:%
\[
\phi\left(  y\right)  =Ne^{-\frac{\left(  y-y_{0}\right)  ^{2}}{4\sigma^{2}}%
},
\]
where $y\in\mathbb{R}$, and $N$ is a normalization constant.

We assume that $t=0$, the measurement interaction begins, and the joint
wavefunction of the particle + pointer system is%
\[
\Psi_{0}\left(  x,y\right)  =\Theta_{0}\left(  x\right)  \phi\left(  y\right)
\text{.}%
\]
The system particle + pointer is described by the state $\Psi\left(
x,y,t\right)  $ satisfying%
\[
i\frac{\partial}{\partial t}\Psi\left(  x,y,t\right)  =\boldsymbol{H}\text{
}\Psi\left(  x,y,t\right)  \text{, \ with }x\in\mathbb{X}\text{, }%
y\in\mathbb{R},t\geq0.
\]
The Hamiltonian $\boldsymbol{H}$ is the sum of three different Hamiltonians.
The first one corresponding to the kinetic \ and potential energies of the
particle in the box, whose degree of freedom is $x$:
\[
\boldsymbol{H}_{x}=\left\{
\begin{array}
[c]{ll}%
-\frac{1}{2m}\frac{\partial^{2}}{\partial x^{2}}+V(x) & \text{if }%
x\in\mathbb{R}\\
& \\
-\frac{1}{2m}\boldsymbol{D}+V(\left\vert x\right\vert _{p}) & \text{if }%
x\in\mathbb{Q}_{p},
\end{array}
\right.
\]
where $\boldsymbol{D}$ is the Vladimirov operator, see \ Appendix C. We assume
that $\boldsymbol{H}_{x}\psi_{n}\left(  x\right)  =E_{n}\psi_{n}\left(
x\right)  $.

The second Hamiltonian comes from the kinetic energy of the pointer whose
degree of freedom is $y$:%
\[
\boldsymbol{H}_{y}=-\frac{1}{2M}\frac{\partial^{2}}{\partial y^{2}},
\]
where $M$ is the mass of the pointer, which we assume to be large, which
warrants $\boldsymbol{H}_{y}\approx\boldsymbol{0}$. The third Hamiltonian
describes the interaction of the particle and the pointer, see \cite[Chapter
3]{Norsen} for a further discussion. We take%
\[
\boldsymbol{H}_{\text{int}}=-\lambda i\boldsymbol{H}_{x}\frac{\partial
}{\partial y},
\]
where $\lambda$ is a constant describing the strength of the interaction. We
assume that $\lambda$ is very large so $\boldsymbol{H}\approx\boldsymbol{H}%
_{\text{int}}$. It is crucial to mention that this approximation does not
depend on eigenfunctions of operator $\boldsymbol{H}_{x}$.

Then, the Schr\"{o}dinger equation of the particle + pointer \ system is
\begin{equation}
i\frac{\partial}{\partial t}\Psi\left(  x,y,t\right)  =\boldsymbol{H}%
_{\text{int}}\text{ }\Psi\left(  x,y,t\right)  =-\lambda i\boldsymbol{H}%
_{x}\frac{\partial}{\partial y}\Psi\left(  x,y,t\right)  .
\label{Eq_Schrodinger}%
\end{equation}
Assuming that $\Psi\left(  x,y,t\right)  =\psi_{n}\left(  x\right)
\Psi_{\bullet}\left(  y,t\right)  $, and using that $\boldsymbol{H}_{x}%
\psi_{n}\left(  x\right)  =E_{n}\psi_{n}\left(  x\right)  $,
(\ref{Eq_Schrodinger}) becomes
\[
\frac{\partial}{\partial t}\Psi_{\bullet}\left(  y,t\right)  =-\lambda
E_{n}\frac{\partial}{\partial y}\Psi_{\bullet}\left(  y,t\right)  .
\]
The general solution of this last equation is
\[
\Psi_{\bullet}\left(  y,t\right)  =\Phi\left(  y-\lambda E_{n}t\right)  ,
\]
where $\Phi$ is any differentiable function. By taking $\Phi\left(  y\right)
=\phi\left(  y\right)  $, and using superposition,%
\[
\Psi\left(  x,y,t\right)  =%
{\displaystyle\sum\limits_{n=1}^{\infty}}
c_{n}\psi_{n}\left(  x\right)  \phi\left(  y-\lambda E_{n}t\right)  ,
\]
where the complex constants $c_{n}$ are determined by the condition
$\Psi\left(  x,y,0\right)  =\Theta_{0}\left(  x\right)  \phi\left(  y\right)
$. Suppose that the interaction lasts until $t=T$. Then, the quantum state of
the particle + pointer system at the end of the interaction is%
\[
\Psi\left(  x,y,T\right)  =%
{\displaystyle\sum\limits_{n=1}^{\infty}}
c_{n}\psi_{n}\left(  x\right)  \phi\left(  y-\lambda E_{n}T\right)  .
\]
The interpretation of this result is that the particle in the box does not end
up in a particular energy eigenstate at all, and worse, the pointer is not
localized around any particular one of its possible final positions. See
\cite[Chapter 3]{Norsen}, for an in-depth discussion of the standard case.

\subsection{The $p$-adic model for a particle in a box}

We use the results from \cite{Zuniga-Mayes} for dimension one, explicitly we
use the following parameters $N=1$, $\alpha=1$, $m_{\alpha}=1$, $L=0$, and
$p\geq3$.

The Schr\"{o}dinger equation for a \ particle in a box is
\begin{equation}
\left\{
\begin{array}
[c]{ll}%
i\frac{\partial\Psi(x,t)}{\partial t}=\boldsymbol{D}\Psi(x,t)\text{,} &
x\in\mathbb{Z}_{p},\quad t\geq0\\
& \\
\Psi(x,t)=0, & x\notin\mathbb{Z}_{p},\quad t\geq0,
\end{array}
\right.  \label{Schrodinger_Equation_1}%
\end{equation}
where the box is the unit ball, and the potential is
\[
V(x)=V(\left\vert x\right\vert _{p})=\left\{
\begin{array}
[c]{ccc}%
0 & \text{if} & x\in\mathbb{Z}_{p}\\
&  & \\
\infty & \text{if} & x\notin\mathbb{Z}_{p},
\end{array}
\right.
\]
We look for solutions of the time-dependent Schr\"{o}dinger equation
(\ref{Schrodinger_Equation_1}) of the form
\[
\Psi(x,t)=e^{-iEt}\Phi(x),
\]
where $\Phi(x)$ is the time-independent function satisfying
\begin{equation}
\left\{
\begin{array}
[c]{lll}%
\Phi(x)\in L^{2}(\mathbb{Z}_{p}) &  & \\
&  & \\
\boldsymbol{D}\Phi(x)=E\Phi(x)\text{,} & \text{for } & x\in\mathbb{Z}_{p}\\
&  & \\
\Phi(x)=0 & \text{for } & x\notin\mathbb{Z}_{p}.
\end{array}
\right.  \text{ } \label{Schrodinger_Equation_Ind_0}%
\end{equation}
Without loss of generality, \ we may assume that $\Phi(x)$ is a real-valued
function. The solution of this last eigenvalue problem is given in
\cite[Theorem 7.1]{Zuniga-Mayes}: the solutions to the \ eigenvalue problem%
\[
\left\{
\begin{array}
[c]{l}%
\Phi_{E}(x)\in L_{\mathbb{R}}^{2}\left(  \mathbb{Z}_{p}\right)  ;\text{
}\left\Vert \Phi_{E}\right\Vert _{2}=1\\
\\
\boldsymbol{D}\Phi_{E}(x)=E\Phi_{E}(x)
\end{array}
\right.
\]
have the following form: for $E=E_{gnd}:=\frac{\left(  1-p^{-1}\right)
}{\left(  1-p^{-2}\right)  }=\frac{1}{1+p^{-1}}$, $\Phi_{E_{gnd}}%
(x)=\Omega\left(  \left\vert x\right\vert _{p}\right)  $; and for
$E_{r}=p^{(1-r)}$, with $r\leq0$,
\begin{equation}
\Phi_{E_{r}}(x)=2p^{\frac{-r}{2}}\Omega(\left\vert p^{r}x\right\vert _{p})%
{\displaystyle\sum\limits_{k\in\mathbb{H}_{p}^{+}}}
A_{k}\cos{\left(  2\pi\{p^{r-1}kx\}_{p}\right)  }, \label{Formula_Phi}%
\end{equation}
where the $A_{k}\in\mathbb{R}$ satisfy
\[
\sqrt{%
{\displaystyle\sum\limits_{k\in\mathbb{H}_{p}^{+}}}
A_{k}^{2}}=\frac{1}{\sqrt{2}},
\]
where $\mathbb{H}_{p}^{+}=\left\{  1,2,\ldots,\frac{p-1}{2}\right\}  .$

Here, it is more convenient to change the notation by taking $-r=l\geq0$, and
$\Phi_{l}(x)=\Phi_{E_{r}}(x)$, $l\geq0$, then%
\begin{equation}
\Phi_{l}(x)=2p^{\frac{l}{2}}\Omega(\left\vert p^{-l}x\right\vert _{p})%
{\displaystyle\sum\limits_{k\in\mathbb{H}_{p}^{+}}}
A_{k}\cos{\left(  2\pi\{p^{-l-1}kx\}_{p}\right)  }, \label{Formula_Phi-2}%
\end{equation}
and $\Phi_{gnd}(x)=\Omega\left(  \left\vert x\right\vert _{p}\right)  $. The
energy levels are $\left\{  \frac{1}{1+p^{-1}},p,p^{2},\ldots,p^{n}%
,\ldots\right\}  $.

\subsection{The Monna map}

The Monna map is defined as%
\[%
\begin{array}
[c]{cccc}%
\mathcal{M}: & \mathbb{Q}_{p} & \rightarrow & \mathbb{R}_{\geq0}\\
&  &  & \\
& x_{p}=%
{\displaystyle\sum\limits_{j=\gamma}^{\infty}}
y_{j}p^{j} & \rightarrow & x_{\infty}=%
{\displaystyle\sum\limits_{j=\gamma}^{\infty}}
y_{j}p^{-j-1}.
\end{array}
\]
This map captures the strangeness of QM: $\mathbb{R}$ does not contain a copy
of $\mathbb{Q}_{p}$ that preserves both the topology and algebraic structure
of $\mathbb{Q}_{p}$. The Monna map is a continuous, surjective, but not
injective, \cite[Section 1.9.4]{Alberio et al}. We use the identity%
\[
\mathcal{M}\left(  p^{l}a+p^{l}\mathbb{Z}_{p}\right)  =\mathcal{M}%
(p^{l}a)+\left[  0,p^{-l}\right]  =\left[  \mathcal{M}(p^{l}a),\text{
}\mathcal{M}(p^{l}a)+p^{-l}\right]  .
\]
Taking $a=p^{-1}$, we have $\mathcal{M}\left(  p^{l-1}+p^{l}\mathbb{Z}%
_{p}\right)  =p^{-l}+\left[  0,p^{-l}\right]  $.

\subsection{The collapse in a discrete space}

We now\ take $\mathbb{X}=\mathbb{Q}_{p}$, \ then $\boldsymbol{H}_{x}=$
$-\frac{1}{2m}\boldsymbol{D}+V(\left\vert x\right\vert _{p})$, with
$x\in\mathbb{Q}_{p}$. Then the wavefunction of the particle + pointer system
at $t=T$ is%

\begin{equation}
\Psi\left(  x,y,T\right)  =A\Phi_{gnd}(x)\phi\left(  y-\frac{\lambda}%
{1+p^{-1}}T\right)  +%
{\displaystyle\sum\limits_{l=0}^{\infty}}
c_{l}\Phi_{l}(x)\phi\left(  y-\lambda p^{1+l}T\right)  ,
\label{Expansion_Spsi}%
\end{equation}
where $A$, and the $c_{l}$ are complex constants.

We assume that during the measurement process the apparatus scans a small
region of $\mathbb{R}$ corresponding to the\ interval $p^{-n}+\left[
0,p^{-n}\right]  $, $n\geq1$, which corresponds to the ball $p^{n-1}%
+p^{n}\mathbb{Z}_{p}$ by the Monna map: $\mathcal{M}\left(  p^{n-1}%
+p^{n}\mathbb{Z}_{p}\right)  =\left[  p^{-n},2p^{-n}\right]  $. This means
that in (\ref{Expansion_Spsi}), $y\in\left[  p^{-n},2p^{-n}\right]
\subset\mathbb{R}$, and $x\in p^{n-1}+p^{n}\mathbb{Z}_{p}\subset\mathbb{Q}%
_{p}$. Now, using that $x\in p^{n-1}+p^{n}\mathbb{Z}_{p}\Leftrightarrow$
$\left\vert x\right\vert _{p}=p^{-n+1}$, and $\Omega(\left\vert p^{-l}%
x\right\vert _{p})=1\Leftrightarrow\left\vert x\right\vert _{p}\leq p^{-l}$,
we conclude
\[
\Omega(\left\vert p^{-l}x\right\vert _{p})=\left\{
\begin{array}
[c]{ll}%
1 & \text{if }l+1\leq n\\
& \\
0 & \text{otherwise;}%
\end{array}
\right.
\]
which implies that%
\[
\Psi\left(  x,y,T\right)  =A\phi\left(  y-\frac{\lambda}{1+p^{-1}}T\right)  +%
{\displaystyle\sum\limits_{l=0}^{n-1}}
c_{l}\left\{  2p^{\frac{l}{2}}%
{\displaystyle\sum\limits_{k\in\mathbb{H}_{p}^{+}}}
A_{k}\cos{\left(  2\pi\{p^{-l-1}kx\}_{p}\right)  }\right\}  \phi\left(
y-\lambda p^{1+l}T\right)  .
\]
Now, if $x\in p^{n-1}+p^{n}\mathbb{Z}_{p}$ and $l\leq n-2$, then
$x=p^{n-1}+p^{n}\widetilde{x}$, with $\widetilde{x}\in\mathbb{Z}_{p}$, and%
\[
p^{-l-1}kx=p^{n-l-2}k+p^{n-l-1}k\widetilde{x}\in\mathbb{Z}_{p}\text{,}%
\]
consequently%
\[
\cos{\left(  2\pi\{p^{-l-1}kx\}_{p}\right)  =1}\text{{, for }}l\leq n-2.
\]
Then, $\Psi\left(  x,y,T\right)  $ can be rewritten as%
\begin{multline*}
\Psi\left(  x,y,T\right)  =C\phi\left(  y-\frac{\lambda}{1+p^{-1}}T\right)  +%
{\displaystyle\sum\limits_{l=0}^{n-2}}
C_{l}\phi\left(  y-\lambda p^{1+l}T\right)  +\\
C_{n-1}\left\{
{\displaystyle\sum\limits_{k\in\mathbb{H}_{p}^{+}}}
A_{k}\cos{\left(  2\pi\{p^{-n}kx\}_{p}\right)  }\right\}  \phi\left(
y-\lambda p^{n}T\right)  .
\end{multline*}
We now take $\phi\left(  y\right)  =Ne^{-\frac{\left(  y-y_{0}\right)  ^{2}%
}{4\sigma^{2}}}$, and interpret $x$, and $T$ as parameters, so $\Psi\left(
x,y,T\right)  =\Psi\left(  y\right)  $, and
\begin{multline*}
\Psi\left(  y\right)  =De^{-\frac{\left(  y-\frac{\lambda}{1+p^{-1}}%
T-y_{0}\right)  ^{2}}{4\sigma^{2}}}+%
{\displaystyle\sum\limits_{l=0}^{n-2}}
D_{l}e^{-\frac{\left(  y-\lambda p^{1+l}T-y_{0}\right)  ^{2}}{4\sigma^{2}}}+\\
D_{n-1}\Omega\left(  x,n\right)  e^{-\frac{\left(  y-\lambda p^{n}%
T-y_{0}\right)  ^{2}}{4\sigma^{2}}},\text{ }y\in\left[  p^{-n},2p^{-n}\right]
\end{multline*}
where $x\in p^{n-1}+p^{n}\mathbb{Z}_{p}$, $D$, $D_{l}$ are constants, and%
\[
\Omega\left(  x,n\right)  :=\sum\limits_{k\in\mathbb{H}_{p}^{+}}A_{k}%
\cos{\left(  2\pi\{p^{-n}kx\}_{p}\right)  }.
\]
For $x=p^{n-1}+p^{n}\widetilde{x}$, with $\widetilde{x}\in\mathbb{Z}_{p}$, we
have $\{p^{-n}kx\}_{p}=\frac{k}{p}$, and thus%
\[
\Omega\left(  x,n\right)  =\sum_{k\in\mathbb{H}_{p}^{+}}A_{k}\cos\left(
\frac{2\pi k}{p}\right)
\]
does not depend on $x$; we assume that this number is different from zero.
This means that the wavefunction is localized in the macroscopic space.

We now assume that the purpose of the interaction is to measure the energy
level $p^{n}$, and take $y_{0}=-\lambda p^{n}T$. Then the last term in
$\Psi\left(  y\right)  $ becomes $D_{n-1}\Omega\left(  x,n\right)
e^{-\frac{y^{2}}{4\sigma^{2}}}$. For $0\leq l\leq n-2$,
\[
y-\lambda p^{1+l}T-y_{0}=y+\lambda p^{n}T\left(  1-p^{l+1-n}\right)
\geq\lambda p^{n}T\left(  1-p^{-1}\right)  ,
\]
because $y>0$ and $1-p^{l+1-n}\geq1-p^{-1}$; the same lower bound holds for
the term corresponding to the ground state. Therefore, for $y\in\left[
p^{-n},2p^{-n}\right]  $, all the terms except the last one are bounded by
$\exp\left(  -\frac{\lambda^{2}p^{2n}T^{2}\left(  1-p^{-1}\right)  ^{2}%
}{4\sigma^{2}}\right)  $, which is negligible in the strong coupling regime
$\frac{\lambda p^{n}T}{\sigma}\gg1$ (in particular, when $\sigma\approx T$ and
$\lambda$ is large, as assumed above). Consequently,
\[
\Psi\left(  y\right)  \approx D_{n-1}\Omega\left(  x,n\right)  e^{-\frac
{y^{2}}{4\sigma^{2}}}=\mathcal{N}e^{-\frac{y^{2}}{4\sigma^{2}}}\text{ for
}y\in\left[  p^{-n},2p^{-n}\right]  ,
\]
see the last subsection of this appendix for precise estimates. Since
$\Omega\left(  x,n\right)  $ does not depend on $x$ in the ball $p^{n-1}%
+p^{n}\mathbb{Z}_{p}$, the reading does not depend on the position of the
particle inside the scanned ball. Notice also that the ball $p^{n-1}%
+p^{n}\mathbb{Z}_{p}$ does not meet the supports of the eigenfunctions
$\Phi_{l}$ with $l\geq n$, while the eigenfunctions with $l\leq n-1$ are
constant on it; in the notation of (\ref{Table-B}), $\mathcal{N}_{1}%
=\emptyset$. In conclusion, in the space-time $\mathbb{R}\times\left(
\mathbb{R}\times\mathbb{Q}_{p}\right)  $, the Schr\"{o}dinger equation
controls the evolution of the particle + pointer system, and the localization
produces a definite reading without the collapse postulate. Passing from
$\mathbb{R}$ to $\mathbb{R}\times\mathbb{Q}_{p}$ requires an extra,
non-Archimedean, coordinate; notice that $\mathbb{Q}_{p}$ has topological
dimension zero, so this is an extra degree of freedom rather than an extra
dimension in the topological sense.

\subsection{Estimates}

In this subsection, we give the estimates used in the subsection ``The
collapse in a discrete space''. We keep the notation of that subsection:
$p\geq3$, $n\geq1$, $T>0$, $\lambda>0$, $\sigma>0$, $y_{0}=-\lambda p^{n}T$,
and $x\in p^{n-1}+p^{n}\mathbb{Z}_{p}$, i.e., $x=p^{n-1}+p^{n}\widetilde{x}$
with $\widetilde{x}\in\mathbb{Z}_{p}$. Notice that $\left\vert x\right\vert
_{p}=p^{-n+1}$.

\textbf{Step 1: the eigenfunctions on the ball.} The ground state satisfies
$\Phi_{gnd}(x)=\Omega\left(  \left\vert x\right\vert _{p}\right)  =1$, since
$\left\vert x\right\vert _{p}\leq1$. For $l\geq0$, $\Omega(\left\vert
p^{-l}x\right\vert _{p})=1$ if and only if $\left\vert x\right\vert _{p}\leq
p^{-l}$, i.e., $l\leq n-1$; thus $\Phi_{l}(x)=0$ for $l\geq n$. For $0\leq
l\leq n-2$ and $k\in\mathbb{H}_{p}^{+}$,
\[
p^{-l-1}kx=p^{n-l-2}k+p^{n-l-1}k\widetilde{x}\in\mathbb{Z}_{p},
\]
so $\{p^{-l-1}kx\}_{p}=0$ and $\Phi_{l}(x)=2p^{\frac{l}{2}}\sum_{k}A_{k}$. For
$l=n-1$, $p^{-n}kx=\frac{k}{p}+k\widetilde{x}$ with $k\widetilde{x}%
\in\mathbb{Z}_{p}$, so $\{p^{-n}kx\}_{p}=\frac{k}{p}$, and
\[
\Phi_{n-1}(x)=2p^{\frac{n-1}{2}}\Omega\left(  x,n\right)  ,\qquad\Omega\left(
x,n\right)  =\sum_{k\in\mathbb{H}_{p}^{+}}A_{k}\cos\left(  \frac{2\pi k}%
{p}\right)  .
\]
In particular, every eigenfunction is either constant on the ball
$p^{n-1}+p^{n}\mathbb{Z}_{p}$ or vanishes on it; in the notation of
(\ref{Table-B}), $\mathcal{N}_{1}=\emptyset$.

\textbf{Step 2: the pointer wavefunction.} By Step 1, for $x\in p^{n-1}%
+p^{n}\mathbb{Z}_{p}$,
\[
\Psi\left(  y\right)  =De^{-\frac{\left(  y-\frac{\lambda T}{1+p^{-1}}%
-y_{0}\right)  ^{2}}{4\sigma^{2}}}+\sum_{l=0}^{n-2}D_{l}e^{-\frac{\left(
y-\lambda p^{1+l}T-y_{0}\right)  ^{2}}{4\sigma^{2}}}+D_{n-1}\Omega\left(
x,n\right)  e^{-\frac{y^{2}}{4\sigma^{2}}},
\]
with $D=AN$, $D_{l}=2c_{l}p^{\frac{l}{2}}\left(  \sum_{k}A_{k}\right)  N$, and
$D_{n-1}=2c_{n-1}p^{\frac{n-1}{2}}N$.

\textbf{Step 3: lower bounds for the exponents.} Let $y\in\left[
p^{-n},2p^{-n}\right]  $. For $0\leq l\leq n-2$,
\[
y-\lambda p^{1+l}T-y_{0}=y+\lambda p^{n}T\left(  1-p^{l+1-n}\right)
\geq\lambda p^{n}T\left(  1-p^{-1}\right)  >0.
\]
For the ground state, using $\frac{1}{1+p^{-1}}<1$ and $p^{n}-1\geq
p^{n}\left(  1-p^{-1}\right)  $,
\[
y-\frac{\lambda T}{1+p^{-1}}-y_{0}\geq\lambda T\left(  p^{n}-1\right)
\geq\lambda p^{n}T\left(  1-p^{-1}\right)  >0.
\]

\textbf{Step 4: the estimate.} Set $\varepsilon:=\exp\left(  -\frac
{\lambda^{2}p^{2n}T^{2}\left(  1-p^{-1}\right)  ^{2}}{4\sigma^{2}}\right)  $.
By Step 3, for $y\in\left[  p^{-n},2p^{-n}\right]  $,
\[
\left\vert \Psi\left(  y\right)  -D_{n-1}\Omega\left(  x,n\right)
e^{-\frac{y^{2}}{4\sigma^{2}}}\right\vert \leq\left(  \left\vert D\right\vert
+\sum_{l=0}^{n-2}\left\vert D_{l}\right\vert \right)  \varepsilon.
\]
On the other hand, $e^{-\frac{y^{2}}{4\sigma^{2}}}\geq e^{-\frac{p^{-2n}%
}{\sigma^{2}}}$ on this interval. Therefore, if $D_{n-1}\Omega\left(
x,n\right)  \neq0$,
\[
\sup_{y\in\left[  p^{-n},2p^{-n}\right]  }\frac{\left\vert \Psi\left(
y\right)  -D_{n-1}\Omega\left(  x,n\right)  e^{-\frac{y^{2}}{4\sigma^{2}}%
}\right\vert }{\left\vert D_{n-1}\Omega\left(  x,n\right)  \right\vert
e^{-\frac{y^{2}}{4\sigma^{2}}}}\leq\frac{\left\vert D\right\vert +\sum
_{l=0}^{n-2}\left\vert D_{l}\right\vert }{\left\vert D_{n-1}\Omega\left(
x,n\right)  \right\vert }\,e^{\frac{p^{-2n}}{\sigma^{2}}}\,\varepsilon,
\]
which tends to zero as $\frac{\lambda p^{n}T}{\sigma}\rightarrow\infty$, for
fixed coefficients and fixed $\sigma$ bounded away from zero. In particular,
with $\sigma\approx T$, the relative error is of order $\exp\left(
-\frac{\lambda^{2}p^{2n}\left(  1-p^{-1}\right)  ^{2}}{4}\right)  $, which is
negligible for large $\lambda$. This justifies the approximation $\Psi\left(
y\right)  \approx\mathcal{N}e^{-\frac{y^{2}}{4\sigma^{2}}}$ on $\left[
p^{-n},2p^{-n}\right]  $ used in the subsection ``The collapse in a discrete space''.

\section{\label{Appendix E}Appendix E: Embedding of finite-dimensional
scenarios}

\subsection{Embedding of multipartite finite-dimensional QM}

Let $\mathbb{C}^{N_{1}}\otimes\cdots\otimes\mathbb{C}^{N_{k}}$ be the Hilbert
space of a finite-dimensional scenario with $k$ systems (for instance, the two
qubits and the two Friends' memories in \cite{Bong et al}). For each $j$, fix
$l_{j}$ with $N_{j}\leq p^{l_{j}}$ and, as in Appendix B, an isometry%
\[
V_{j}:\mathbb{C}^{N_{j}}\rightarrow L^{2}(\mathbb{Z}_{p}),\qquad e_{I}\mapsto
p^{\frac{l_{j}}{2}}\Omega\left(  p^{l_{j}}\left\vert x-I\right\vert
_{p}\right)  ,\quad I\in G_{l_{j}}^{0},
\]
with range $\chi_{N_{j}}(\mathbb{Z}_{p})$. Set $V=V_{1}\otimes\cdots\otimes
V_{k}:\bigotimes_{j}\mathbb{C}^{N_{j}}\rightarrow\bigotimes_{j}L^{2}%
(\mathbb{Z}_{p})=L^{2}(\mathbb{Z}_{p}^{k})$; $V$ is an isometry with range
$\chi:=\bigotimes_{j}\chi_{N_{j}}(\mathbb{Z}_{p})$, and $VV^{\ast}$ is the
orthogonal projection onto $\chi$.

For an operator $A$ on $\bigotimes_{j}\mathbb{C}^{N_{j}}$, set $\widehat
{A}:=VAV^{\ast}$. Then: (i) if $H$ is Hermitian, $\widehat{H}$ is a bounded
self-adjoint operator on $L^{2}(\mathbb{Z}_{p}^{k})$, which vanishes on
$\chi^{\perp}$; for $k=1$, $\widehat{H}$ is the integral operator
(\ref{Integral}) of Appendix B; (ii) $e^{-it\widehat{H}}=Ve^{-itH}V^{\ast
}+\left(  1-VV^{\ast}\right)  $; (iii) if $P$ is an orthogonal projection,
$\widehat{P}$ is an orthogonal projection on $L^{2}(\mathbb{Z}_{p}^{k})$; (iv)
if $A=A_{1}\otimes1\otimes\cdots\otimes1$ acts on the first factor, then
$\widehat{A}=V_{1}A_{1}V_{1}^{\ast}\otimes V_{2}V_{2}^{\ast}\otimes
\cdots\otimes V_{k}V_{k}^{\ast}$, and, in particular, $\widehat{A}$ commutes
with every $\widehat{B}$, where $B$ acts on the other factors. Properties
(i)--(iii) follow from $V^{\ast}V=1$; (ii) follows from (i) by the functional calculus.

Let $\psi$ be a state and let $U_{1},\ldots,U_{s}$ be unitary operators and
$P_{1},\ldots,P_{s}$ orthogonal projections on $\bigotimes_{j}\mathbb{C}%
^{N_{j}}$. Since $\chi$ is invariant under all the operators $\widehat{U}_{i}%
$, $\widehat{P}_{i}$,%
\[
\left\Vert \widehat{P}_{s}\widehat{U}_{s}\cdots\widehat{P}_{1}\widehat{U}%
_{1}V\psi\right\Vert ^{2}=\left\Vert P_{s}U_{s}\cdots P_{1}U_{1}%
\psi\right\Vert ^{2}.
\]
Thus all the joint probabilities of sequences of unitary evolutions and
projective measurements coincide with those of standard QM. In the scenario of
\cite{Bong et al}, the Friends' measurements are unitary interactions, and the
measurements of Alice and Bob are projective; therefore, the Local
Friendliness inequalities are violated by the amount predicted by QM. Since
the operations of Alice and Bob act on different tensor factors, the marginal
distribution of the outcomes of Alice does not depend on the settings of Bob,
and vice versa, exactly as in standard QM.

\section{\label{Appendix F}Appendix F: The role of the update rule}

In this appendix, we analyze the alternative reading of the localization
mentioned in Subsection \ref{Section_Further}, in which the projection
(\ref{Update_rule}) does not change the state, and we show what the update
rule adds to the framework.

\subsection{The framework without update}

Suppose that the global state of all the systems involved evolves according to
the Schr\"{o}dinger equation at all times and is never replaced by
(\ref{Update_rule}), (\ref{Lift}). The reading of an apparatus is still
determined by the projection (\ref{Update_rule}) of the global state onto the
graph of $\mathcal{M}$ over a ball $B$, with $B$ selected with probability
$P_{\text{int}}(B,T)$. The results of Subsections \ref{Section_Meas_I} and
\ref{Section_Meas_II} apply without change: each reading is definite, since
the collapse mechanism does not depend on the update rule. What changes is the
initial condition for the evolution after each measurement. The difference
concerns sequences of measurements, since the global state at the time of
Wigner's measurement contains the components corresponding to all the possible
readings of the Friend. In the rest of this appendix, we consider only the
sequence of Scenario I of Section \ref{Section_4}: the Friend's measurement
followed by Wigner's measurement, $T_{\text{Friend}}<T_{\text{Wigner}}$; longer
sequences of measurements are not studied here.

\subsection{Absence of interference terms}

Let $B^{(1)},\ldots,B^{(s)}$ and the components $\Psi_{j}$ be as in Subsection
\ref{Section_Persist}, and let $\Psi$ be the global state at
$T_{\text{Friend}}$, which we assume supported, in $x_{p}$, in the union of the
balls $B^{(j)}$, so that $\Psi=\sum_{j}\Psi_{j}$. Assume that the Hamiltonian $\boldsymbol{H}_{W}$
governing $\mathcal{S}_{1}+\mathcal{W}$ between $T_{\text{Friend}}$ and
$T_{\text{Wigner}}$ commutes with the multiplication operators $1_{B^{(j)}%
}(x_{p})$. Then $e^{-it\boldsymbol{H}_{W}}$ also commutes with them, and
$e^{-it\boldsymbol{H}_{W}}\Psi_{j}$ is supported in $B^{(j)}$, in the variable
$x_{p}$, for all $t$. Let $B_{W}$ be the ball selected in Wigner's measurement, of radius
$p^{-l_{W}}$, and let $G$ be as
in Subsection \ref{Section_Meas_II}. Since the functions $\Psi_{j}%
(T_{\text{Wigner}},G(x_{p}))$ have disjoint supports in $x_{p}$,%
\[
\int\limits_{B_{W}}\left\vert \sum_{j}\Psi_{j}\left(  T_{\text{Wigner}%
},G(x_{p})\right)  \right\vert ^{2}dx_{p}=\sum_{j}\int\limits_{B_{W}\cap
B^{(j)}}\left\vert \Psi_{j}\left(  T_{\text{Wigner}},G(x_{p})\right)
\right\vert ^{2}dx_{p}.
\]
There are no interference terms: Wigner's probabilities are those of a mixture
of the Friend's readings, and the joint probability of the Friend's reading
$B^{(j)}$ and Wigner's reading $B_{W}$ is well defined,%
\begin{equation}
P\left(  B^{(j)},B_{W}\right)  =\frac{1}{C\left(  \mathcal{M},\mathcal{M}_{\ast
}\right)  }\int\limits_{B^{(j)}\cap B_{W}}\left\vert \Psi\left(
T_{\text{Wigner}},G(x_{p})\right)  \right\vert ^{2}dx_{p}. \label{Joint_prob}%
\end{equation}
Recall that a ball of radius $p^{-l}$ fixes the first $l$ digits of $x_{p}$.
By the ultrametric property, two balls are either disjoint or nested, so one of
the following holds. If $l_{W}\geq l_{F}$, then either $B_{W}\subseteq
B^{(j)}$ for exactly one $j$, so that Wigner's reading is finer than the
Friend's and determines it, or $B_{W}$ is disjoint from all the $B^{(j)}$. If
$l_{W}<l_{F}$, then $B_{W}$ contains the balls $B^{(j)}$ that it meets, so that
Wigner's reading is coarser than the Friend's and is determined by it.

The assumption on $\boldsymbol{H}_{W}$ means that the evolution between the
two measurements does not move amplitude from one ball $B^{(j)}$ to another.
Operators acting only on $x_{\infty}$ and $y_{\infty}$, and operators of
multiplication by functions of $x_{p}$, commute with the operators
$1_{B^{(j)}}(x_{p})$, so a Hamiltonian built from them satisfies the
assumption. The Vladimirov operator acting on $x_{p}$ does not commute with
them: it is non-local, and it moves amplitude from one ball to another; for
instance, $\boldsymbol{D}^{\alpha}1_{\mathbb{Z}_{p}}(x)\neq0$ for $\left\vert
x\right\vert _{p}>1$ (Appendix C). Thus, even without the update rule, there
are no interference terms between the Friend's readings in Wigner's
statistics, provided that the dynamics between the two measurements does not
move amplitude between the balls $B^{(j)}$; the ultrametric property then
gives the simple relation between the two readings described above.

The case $B_{W}\supseteq B^{(j)}$ does not mean that Wigner undoes the
Friend's record. Under the assumption on $\boldsymbol{H}_{W}$, the components
$\Psi_{j}$ remain supported in the disjoint balls $B^{(j)}$, so the record of
the Friend's reading persists; Wigner's measurement only fails to resolve it.
Indeed, the probability of Wigner's reading is
\[
P\left(  B_{W}\right)  =\sum_{j:\,B^{(j)}\subseteq B_{W}}P\left(
B^{(j)},B_{W}\right)  ,
\]
a sum without interference terms, as for a coarse measurement of a mixture of
the Friend's readings. Without the update rule, a record of the Friend's reading
$B^{(j)}$ is undone, in the sense of Subsection \ref{Section_Persist}, when the
evolution moves amplitude between the balls and recombines the components
$\Psi_{j}$, so that they are no longer separated by the projections
$1_{B^{(j)}}(x_{p})$. A later measurement can then display interference terms between the Friend's possible
readings, and the probabilities of Wigner's readings are no longer those of a
mixture of the Friend's readings. This requires a Hamiltonian that does not
commute with the operators $1_{B^{(j)}}(x_{p})$, which is excluded by the
assumption of this subsection. For example, if the Friend's measurement maps $(\alpha\left\vert
\uparrow\right\rangle +\beta\left\vert \downarrow\right\rangle )\otimes
\left\vert \text{ready}\right\rangle $ to $\alpha\left\vert \uparrow
\right\rangle \left\vert F_{\uparrow}\right\rangle +\beta\left\vert
\downarrow\right\rangle \left\vert F_{\downarrow}\right\rangle $, with the two
branches supported in disjoint balls, the inverse interaction returns the
global state to $(\alpha\left\vert \uparrow\right\rangle +\beta\left\vert
\downarrow\right\rangle )\otimes\left\vert \text{ready}\right\rangle $, and a
subsequent measurement can display interference between the two branches. Such an inverse interaction must act on $x_{p}$ across the
balls, for instance through the Vladimirov operator. The free $p$-adic evolution of the system does
not commute with $1_{B^{(j)}}(x_{p})$ either, so, without the update rule, the
records may also degrade without any intervention of Wigner.

\subsection{Consequences}

Without the update rule, the evolution is unitary for all systems, including
apparatuses, so the premises of the no-go theorems of Section \ref{Section_5}
hold: the Friend's measurement can be reversed, and the Friend's reading is an
observed event. The framework then keeps the absoluteness of each reading and
no-superdeterminism, and must surrender locality, as allowed by \cite{Bong et
al}, and the persistence of records that are undone, as required by
\cite{Guerin et al}. The theorem of Bong et al. then requires a non-local
joint distribution of the readings of the Friends and of the super-observers,
including the readings that are later undone; formula (\ref{Joint_prob}) does
not provide it, since it presupposes that the records persist. Moreover, the
predictions for extended Wigner's Friend experiments coincide with those of
unitary QM. The following table summarizes the comparison.

\begin{center}
{\small
\begin{tabular}
[c]{|l|c|c|}\hline
Feature & With update rule & Without update rule\\\hline
Global evolution & Schr\"{o}dinger + update & unitary at all times\\\hline
Interference of Friend's readings & excluded & excluded if $[\boldsymbol{H}%
,1_{B^{(j)}}]=0$\\\hline
Persistence of records & always & condition on the dynamics\\\hline
Assumptions surrendered & unitarity for apparatuses & locality,
persistence\\\hline
Joint distribution of readings & given by the update & open if records are
undone\\\hline
Local Friendliness violations &  & \\
for apparatus Friends & cease & as in unitary QM\\\hline
\end{tabular}
}
\end{center}

In conclusion, the ultrametric geometry alone makes each reading definite and,
under a condition on the dynamics, excludes interference between successive
readings. The update rule (\ref{Update_rule}) is what guarantees the
persistence of records for every dynamics, determines the joint statistics of
successive readings, and yields the testable prediction of Subsection
\ref{Section_Exp}.

\end{document}